\documentclass[12pt]{aastex}
\usepackage{wasysym}

\newcommand{\msb}{M$_{\odot}$~}

\title{Deep Mixing in Evolved Stars:\\
I. The Effect of Reaction Rate Revisions from C to Al.}

\author{S. Palmerini\altaffilmark{1}, M. La Cognata\altaffilmark{2}, S. Cristallo\altaffilmark{3},
\& M. Busso\altaffilmark{1}}

\altaffiltext{1}{Dipartimento di Fisica, Universit\`{a} di Perugia, and INFN, Sezione di Perugia;
sara.palmerini@fisica.unipg.it}
\altaffiltext{2}{Dipartimento di Metodologie Fisiche e Chimiche per l'Ingegneria,
Universit\`{a} di Catania, Catania, Italy and Laboratori Nazionali del Sud -
INFN, Catania, Italy}
\altaffiltext{3}{Departamento de Fisica Teorica y del Cosmos, Universidad de Granada, Spain}
\begin{document}
\begin{abstract}
We present computations of nucleosynthesis in low-mass
red-giant-branch and asymptotic-giant-branch  stars of Population I
experiencing extended mixing. We adopt the updated version of the
FRANEC evolutionary model, a new post-process code for non-convective mixing and
the most recent revisions for solar abundances. In this framework,
we discuss the effects of recent improvements in relevant reaction
rates for proton captures on intermediate-mass nuclei (from carbon
to aluminum). For each nucleus we briefly discuss the new choices
and their motivations. The calculations are then performed on the
basis of a parameterized circulation, where the effects of the new
nuclear inputs are best compared to previous works. We find that the
new rates (and notably the one for the
$^{14}$N(p,$\gamma$)$^{15}$O reaction) imply considerable
modifications in the composition of post-main sequence stars. In
particular, the slight temperature changes due to the reduced
efficiency of proton captures on $^{14}$N induce abundance variations at
the first dredge up (especially for $^{17}$O, whose equilibrium
ratio to $^{16}$O is very sensitive to the temperature). In this new
scenario presolar oxide grains of AGB origin turn out to be produced
almost exclusively by very-low mass stars ($M \le 1.5 - 1.7
M_{\odot}$), never becoming C-rich. The whole population of
grains with $^{18}$O/$^{16}$O below 0.0015 (the limit permitted by first
dredge up) is now explained. Also, there is now no forbidden area for very low
values of $^{17}$O/$^{16}$O (below 0.0005), contrary to previous findings.
A rather shallow type of transport seems to be sufficient for the CNO changes in
RGB stages. Both thermohaline diffusion and magnetic-buoyancy-induced mixing
might provide a suitable physical mechanism for this.
Thermohaline mixing is in any case certainly
inadequate to account for the production of $^{26}$Al on the AGB. Other
transport mechanisms must therefore  be at play. In general, observational constraints
from RGB and AGB stars, as well as from presolar grains, are well reproduced by
our approach. An exception remains the nitrogen isotopic ratio in mainstream SiC
grains. For the low values measured in them (i.e. for $^{14}$N/$^{15}$N $\le$
2000) we have no explanation. Actually, for the several grains with subsolar
nitrogen isotopic ratios no known stellar process acting in low mass stars can
provide a clue. This might be an evidence that some form of contamination from
cosmic ray spallation occurs in the interstellar medium, adding fresh $^{15}$N to
the grains.

\end{abstract}

\section{Introduction}

It is now well known that the processes of astration and gas return
to the Interstellar Medium (ISM)  of galaxies is dominated by low
mass (LM) and intermediate mass (IM) stars, i.e. by objects below
$M$ = 7-8 \msb \citep{sed}. These stars share the properties of
ascending, after the Main Sequence, along the Red Giant Branch (RGB)
and then, after the occurrence of core-He burning, of terminating
their evolution through the so-called Asymptotic Giant Branch (AGB)
stage \citep[see e.g.][]{ir83}. Here they lose mass efficiently
thanks to radial pulsations and to radiation pressure on dust grains
\citep{hw5,gua6, gb8}, powering stellar winds that are enriched with
the products of the internal nucleosynthesis. Tiny solid particles
that are abundant in such winds maintain the signature of those
nuclear processes; samples of the most refractory among those grains
can be recovered from pristine meteorites \citep{zin1} and this fact
has enormously increased the importance of AGB stars as laboratories
for the study of H- and He-burning nucleosynthesis \citep{bgw}.

At the end of the AGB stage, the residual envelopes of LM and IM
stars are lost through the fast ejection of hot, ionized materials
("superwind");  the circumstellar envelope, heated and illuminated
by the now hot central star, forms a planetary nebula, while the
dense, electron-degenerate core starts a blue-ward path, which will
ultimately give birth to a white dwarf \citep{vw3,pm4}. Stars
crossing these evolutionary stages not only represent a fundamental
component of the baryonic mass in galaxies, but also provide us with
invaluable information on the physical and chemical evolution of the
universe. As a consequence, the understanding of moderately massive
stars is a necessary condition for addressing a large part of modern
astrophysical problems.

Unfortunately, embarrassing gaps exist in our knowledge of their
physics,  and fundamental problems affecting their evolution are not
properly treated in common stellar modeling. This is not surprising,
as stellar models provide only a basic description, where important
physical mechanisms (like convection) are oversimplified and others
(e.g. rotation) are often neglected \citep[see e.g.][for a recent
outline of the importance of rotation]{cl10}. One big problem
affecting low-mass-star evolutionary models is revealed by the
isotopic ratios of CNO elements, as observed in stellar photospheres
\citep{harris,cot,gb91,shet,smil9} and in presolar grains of AGB
origin, preserved in meteorites \citep[see][]{nit97,choi,am01}.

Some chemical peculiarities concerning light elements are found very
early in stellar evolution \citep{r5,m9}. Others appear slightly
after the occurrence of the first dredge up (hereafter FDU), where stars populate
the so-called {\it bump} of the luminosity function \citep{c04}.
Then the appearance of anomalies in the composition continues along
the AGB phase \citep{w5,bus10}.  The observational basis for these
issues has grown considerably over the years, confirming chemical
anomalies at least from Li to O, but sometimes extending up to Mg
and Al, including the production of the unstable isotope $^{26}$Al
\citep[see e.g.][]{p3,c94,gra0,gru2,bus03,zin2,nit08}.

The above evidence suggests that non-convective transport mechanisms
occur in stars, especially during the RGB and AGB phases, linking
the envelope to regions where proton captures occur
\citep{sm,b94,c95,cdn,nol03}

Secular mixing phenomena are made easy in RGB and
AGB  phases immediately following an episode of convective
dredge-up, because the H-burning shell then advances in homogeneous
regions, so that the natural barrier opposed to mass transport by
the chemical stratification is not present. Mechanisms of this kind
might actually account for several puzzling astrophysical
observations: from the evolution of $^3$He in the ISM
\citep{w5,gg98,ba02} to the complex phenomena of production and
destruction of Li in the Galaxy, including its long-term decline in
main sequence stars \citep{sp72,sz92,z94} and its re-production in
some (rare) red giants \citep{cb,gua9}. Further implications of slow
mixing probably include the anti-correlations displayed by globular
cluster stars, involving intermediate mass elements up to Mg and
appearing already on the Main Sequence \citep{gra01,t01,rc02}.
Indeed, they might have been inherited from LM or IM progenitors,
hosting proton captures in mixing episodes through the AGB phases.

Among the processes involved in deep-mixing phenomena, those
induced by rotation certainly play an important role
\citep{z92,cl10}, despite difficulties in explaining the observed
isotopic ratios \citep{pal}. Actually, a shear layer, at the contact
between the differentially rotating convective envelope and the more
rigid stellar core is a natural expectation \citep[see e.g.][]{kw}.
A rigid-body behavior in the cores of low-mass stars might be linked
to an interplay of magnetic coupling and meridional circulation
\citep{egg}.

Despite formal differences in the various approaches, most models
considered the chemical mixing induced by rotational effects (and
the  associated angular momentum transport) as a diffusive process,
with parameterized diffusion coefficients \citep[see
e.g.][]{den98,dw96,dw00,weis00}. This approach has also been adopted
in modeling massive, stratified stars, e.g. by \citet{mm4}.
Alternatives were presented by \citet{dt}, who studied the effects
of gravitational waves, and by \citet{bus07a,nor08,d09}. These last authors
discussed the effects of magnetic buoyancy induced by a stellar
dynamo. The presence of such a dynamo in red giants finds supports
in some stellar variability observations \citep{and88} and in the
identification of surface magnetic fields in AGB stars \citep{herp}.
Stellar magnetic fields and their instabilities have been also the
object of previous detailed analysis \citep{sp99,sp02}. However, by
far the most popular scenario in recent years has been that of the
so-called {\it thermohaline} diffusion, induced by an inversion in
the molecular weight $\mu$. In H-burning regions this is due to the
activation of the reaction $^3$He+$^3$He $\rightarrow$ $^4$He + 2p
\citep{egg1,egg2,cz}, which locally reduces $\mu$. The
phenomenon is more general than that; it was recognized long time
ago and it can occur in other astrophysical contexts \citep{ss69,uli}.

Diffusive mechanisms (such as thermohaline mixing) and circulation
phenomena (like those induced e.g. by magnetic buoyancy) yield
abundance changes that are sometimes undistinguishable from one
another, sometimes remarkably different, depending on the velocity
of the transport. Verifying whether observations can provide clues
to discriminate among the possible physical processes generating the
mixing is therefore an important issue. In this respect, interesting new elements
have been offered by the recent 2D- and 3D-calculations by \citet{den10a,den10b}.
After presenting our results, we shall comment in some detail on
these elements, especially in comparison with what magnetic buoyancy can
a priori provide.

Before the effects of different, uncertain mixing models can be
addressed,  however, one has preliminarily to ascertain that the
normalizing solar abundances and the nuclear input data be very
reliable. Using all the upgrades appeared in
the recent literature on these issues, we intend to revisit the coupled
processes of proton capture nucleosynthesis and of deep mixing in red
giants, starting from a parameterized analysis (where nuclear and
abundance effects are best compared to previous work). In section 2
we shall present a general discussion of the set of new inputs adopted,
postponing to a dedicated Appendix the details of the reaction rate
choices made. In section 3 we shall then
discuss recent upgrades and  novelties in stellar models and in section 4
we shall briefly illustrate our technique for performing parameterized,
post-process deep-mixing calculations, starting from the results of
the FRANEC evolutionary code. Then in sections 5 and 6 we shall
outline the nucleosynthesis results thus obtained, showing some
comparisons with previous work, with stellar observations and with
presolar grain measurements of isotopic abundances for elements from
C to Al. Comments on recent models for thermohaline mixing in 2 and 3 dimensions
are then added in a dedicated section (section 7). Finally,
some general conclusions are drawn in section 8.

\section{On Recent Revisions in Solar Abundances and Reaction Rates}
A previous parameterized analysis of
extra-mixing in red giants \citep{nol03} adopted solar abundances
from \citet{ag89}, coupled with the general prescriptions for
reaction rates from the compilation by \citet{NACRE}, hereinafter
simply referred to as NACRE. Since then, however, many remarkable
changes have occurred in both sets of input data. For what concerns
solar abundances, we shall in general adopt the compilation by
\citet{as09}, implying a total solar metallicity of $Z = 0.014$.
We underline that not only elemental, but also
isotopic abundances (as updated in that review) are important for
our analysis; this is especially the case for CNO elements. In
particular, we shall make use of the new estimate of the
$^{15}$N/$^{14}$N ratio, as deduced by studies of the Jovian
atmosphere \citep{fou1,ow}; for it we adopt the value derived by the
Composite Infrared Spectrometer onboard the Cassini spacecraft
(2.22$\pm 0.52 \times$10$^{-3}$, a factor 1.5 smaller than the
terrestrial one), which should represent the initial solar-system
ratio \citep{fou2}. For oxygen isotopes we use the values suggested
by \citet{as09}.

Concerning reaction rates, recent extensive reviews have been published,
summarizing the main experimental changes occurred after
the NACRE compilation, which has been the reference in most papers
on low-mass stellar nucleosynthesis over the last decade. In
particular, we shall discuss here (and adopt throughout the paper)
the recommendations by \citet{ADE10} and \citet{ILI10}, hereinafter simply refereed
to as ADE10 and ILI10, respectively. We instead will not analyze
the work by \citet{beard}, which pertains to stellar burning at much
higher temperatures and to nucleosynthesis in high-density
environments. The work by ADE10 represents the follow-up of the
solar-fusion cross-section review by \citet{ADE98} and includes
subsequent measurements for reactions of the pp chain and of CNO
cycles. It mainly focuses on low-energy cross sections, aiming at
extracting the rates for temperatures typical of the solar core
($T_9$\footnote{For the temperature we adopt the notation
where $T_9=$T$/10^9$K.}$\sim 0.015$). Reaction rates are obtained analytically, in
terms of low-energy expansion coefficients of the astrophysical S(E)
factors. The work by ILI10 is instead an upgrade of the previous
review by \citet{ILI01}. In the new work several improvements have
been introduced: (i) a wider range of atomic masses is covered, from
A=14 to A=40, including some CNO reactions; (ii) updated nuclear
physics inputs are considered in the reaction rate evaluation; (iii)
a calculation procedure improved with respect to that of
\citet{ILI01} is adopted. In detail,  a suited probability
distribution is first associated with each nuclear physics quantity
entering the reaction-rate calculation. Then, the reaction rate
itself is computed for a set of input parameter values, which are
randomly taken, each following the appropriate distribution. In this
way, a probability density function is generated for the reaction
rate reflecting the distribution of the input values. This method
has the important consequence that the upper and lower limits are
estimated with a statistically-appropriate approach, so that a
really meaningful value can be attributed to them. Moreover, the
rate is obtained over a wide temperature range $T_9 = 0.01 - 10$,
almost reaching down to the values found at the inner borders of the
convective envelopes in low-mass AGB stars ($T_9 \sim 0.005$).
It is worth remarking that both compilations, ADE10 and ILI10,
account for the recent results obtained using indirect techniques,
such as the Coulomb Dissociation \citep{baur}, the Asymptotic Normalization
Coefficient \citep{akram97} and the Trojan Horse Method \citep{laco10}.
This makes the low-temperature reaction rates more accurate, thanks to the
possibility of extending the measurements down to the astrophysical range of
energies.

The recommended reaction rates from ADE10 and ILI10 differ,
sometimes  remarkably, from NACRE results. This fact (and the search
for the descending astrophysical consequences) is the main
motivation for the present work.

Proton capture reactions on nitrogen and oxygen isotopes are of special importance in our context. In particular, the reaction rate for $^{14}$N(p,$\gamma$)$^{15}$O by ADE10, reduced by 50\% with respect to NACRE, leads to a reduction in the efficiency of the whole CNO cycling, moving the H-burning shell toward inner stellar regions,
where temperature and density values are higher by about 10\% and 25\%, respectively, as compared to previous stellar models. We shall discuss at length the important consequences of these facts on the envelope abundances. As we shall see, also the new data for proton captures on $^{17}$O induce non-negligible effects on the surface composition. By contrast, no appreciable effect derives from the reduction in the rate for $^{15}$N(p,$\gamma$)$^{16}$O; in fact, it modifies the abundances of $^{16}$O, $^{17}$O, $^{18}$O and $^{19}$F only in the regions where the H-burning efficiency is maximum (T$_9 >$ 0.05), which are untouched by deep mixing.

In general, we use the ADE10
compilation for pp-chain and CNO-cycle reactions, while we adopt the
results from ILI10 for heavier nuclei. However, the two compilations
partly overlap in the region of CNO isotopes. In this case, specific choices
are necessary. In order to motivate them, and to illustrate some of
the novelties with respect to NACRE, we present in the Appendix a discussion of the available data for some key reactions and of our choice.

\section{The new reference stellar models}

Although we have certainly achieved a general understanding of the
conditions  under which chemical elements are produced in stellar
interiors, a predictive nucleosynthesis theory based on first
principles is still lacking. A basic missing ingredient in
developing such a theory is a full understanding of the transport
properties of the turbulent, reactive plasma in which the elements
are forged. Within this context, many efforts have been spent in the
past years to study the properties of the final phases of stellar
evolution by means of theoretical models calculated with the
Frascati RAphson-Newton Evolutionary Code \citep[FRANEC: see][]{chi98}. As an
example, the extra-mixing calculations by \citet{nol03}, which
represent the starting point of the present work, were based on
FRANEC stellar structures taken from \citet{stra03}. This last work
presented parameterized theoretical formulae permitting a simple
derivation of the physical characteristics of AGB stars (such as,
for example, the number of thermal pulses, the mass of the
H-exhausted core and of the envelope, the stellar surface
temperature and so on), by interpolating the outputs of
previously-computed full AGB evolutionary models \citep{stra97}.

Calculations presented in this paper are instead based on new
evolutionary stellar sequences. In particular, we have performed
complete runs (from the pre-Main Sequence to the end of the AGB
phase) for a 1.5 M$_{\odot}$ star and a 2.0 M$_{\odot}$ star, with a
metallicity equal to the most recent estimate for the solar one
\citep{cri09a}. The runs have been
performed with an upgraded version of the FRANEC code. Computations
extending up to the RGB tip have also been included for various
other masses in the range 1 - 1.7 M$_{\odot}$. Our calculations do
not yet include rotation, nor the effects of magnetic fields.
However, as compared to older versions of the FRANEC stellar code
\citep{stra97}, the models here adopted represent a remarkable
improvement, including several new or updated ingredients.

Among the main differences with respect to the previous scenario we
recall: i) the use of a full nuclear network, coupled with the
physical evolution; ii) the adoption of an updated solar chemical
distribution \citep[in particular, the code was revised in the last
years using the][compilation]{lod03}.\footnote{In the post process we use the
updated suggestions by \citet{as09}: the small differences between
this set of data and the one used in the stellar code does not
introduce any consequence on the model physics.} We remind that older versions
of FRANEC used instead the abundances by \citet{ag89}, which are
substantially different; iii) the use of
new reaction rates, of which crucial is in particular the new value
for the $^{14}$N(p,$\gamma$)$^{15}$O reaction (see the Appendix for details); in
the FRANEC code this update was originally implemented when the
measurements by the LUNA experiment became available
\citep{LUNA1,LUNA2}; iv) the adoption of new low-temperature
opacities and their modifications during the evolution, when the
envelope chemical admixture is changed by dredge up and becomes
carbon rich \citep{cri07}; v) the choice, in evolved stages, of a
mass loss law different from the one previously used; this relation
is similar to the one by \citet{vw}, but adopts more recent data on the
Period-Luminosity relation of AGB stars \citep[see][for details] {sgc}; vi) a
different treatment of the radiative/convective interfaces. In particular,
instead of using a bare Schwarzschild
criterion, an exponentially-decreasing profile of the convective velocities
was applied at the inner border of the convective envelope \citep{cri09a}.

In any case, for a discussion of the improvements outlined above as items from i) to vi)
and of their motivations see especially \citet{cri09a}. For the scopes of this
work, we want primarily to underline that the adoption of a full
nuclear network (although it increases the computational
time by a factor 15 to 40, depending on the case considered) has proven to be crucial.
Indeed, computing a set of full models, with the evolution of all the abundances
coupled to the structural and thermodynamic evolution, provides bona-fide abundance
templates, on which to verify the outcomes of any post-process
calculation done for specific purposes. These post-process studies
remain necessary whenever many parameterized cases have to be
considered (as is required, e.g., in the present paper). However,
now their results (in terms of predicted abundances) can be
validated carefully on a smaller set of complete FRANEC
calculations. The inclusion of the extended network into the code was
primarily motivated by other reasons. In fact, the usual approach of
stellar models, to consider only the limited set of reactions controlling
the energy generation, becomes completely inadequate whenever the physical
properties of a structure are strongly related to the chemical composition
changes. This is, e.g., the case of the proton ingestion by the convective
shell driven by the first fully developed thermal pulse, which is
found to occur in very metal-poor AGB stars and during the post-AGB
phases at larger metallicities. \citep{cri09b,her10}. More generally,
the code with the full network can now be used to compute the nucleosynthesis
in the He-shell (including $s$-processing). In the context of the present work
verifying the post-process calculations with a (limited) set of checks from
real, full-network models was important for those isotopes whose abundances
are strongly influenced by the TDU. Typical examples of this are nuclei like
$^{22}$Ne (produced in the thermal pulses), $^{23}$Na (that derives from proton
and neutron captures on these same $^{22}$Ne seeds) and $^{26}$Al. This last nuclide
is produced in the innermost H-burning zones, hence sinks into the He-rich
buffer and is subsequently largely destroyed by neutron-captures at high temperature
($T \ge 3\times10^8K$) during the thermal pulses. The surviving part is then dredged up
again to the H-rich layers and can account for $^{26}$Al/$^{27}$Al ratios in the
envelope of up to $5-6\times10^{-3}$, which are then enhanced by the extra production
of $^{26}$Al in deep mixing episodes.

As we shall discuss in next section, the new models imply important
changes  in the surface isotopic composition already on the RGB, at
the occurrence of the FDU. In the final
evolutionary stages, then, the effects are even more substantial,
especially for what concerns the chemical enrichment of the envelope
along the AGB.

\section{Computations of Extended Mixing and Nucleosynthesis}

Starting from the models described in section 3, and also for the
sake of comparison with previous works, we have performed
parametric calculations, using as free parameters the rate of mass
transport ($\dot M$) and the maximum temperature sampled by the
circulating material ($T_P$), as in \citet{nol03}. This last
parameter is best expressed by the difference $\Delta = \log T_H -
\log T_P$, where $T_H$ is the temperature at which the energy from
the H-burning shell is maximum. We shall use this definition of
$\Delta$ throughout this paper; similarly, we shall often express
$\dot M$ in units of 10$^{-6}$ M$_{\odot}$/yr and we shall call this
parameter $\dot M_6$. We present here results mainly for
solar-metallicity models of 1.5 and 2.0 M$_{\odot}$, although some
extra models have been computed for comparisons with observational
data. The post-process calculations have been run in a way similar
to what was done previously, using older releases of the same FRANEC
stellar code \citep{nol03,bus10}. The post-process program is new
and its results have been cross-checked with the ones obtained with
the older code \citep{palm2}.

During RGB stages, we have introduced extra-mixing after the "bump"
of the  luminosity function (BLF), i.e. after the H-burning shell
has erased the chemical discontinuity left behind by the FDU \citep{c96}.
We shall not repeat here a general discussion
of RGB physics, as it can be easily found elsewhere \citep{cb,pal}.
Our procedure in the calculations was illustrated by \citet{palm1}.
We have followed the abundance changes during the transport due to
nucleosynthesis (partial derivatives with respect to time) and to
displacements across the integration grid (partial derivatives with
respect to distance, times the mixing velocity). One can write this
synthetically as:
$$
{{dY_i \over dt}} = {\partial Y_i \over \partial t} + {\dot M}\cdot {{\partial Y_i \over \partial M}}
$$
Here $Y_i$ is the abundance by  mole of the i-th isotope (for the
present work,  this index spans from H to $^{32}$S). In
agreement with \citet{nol03} and with \citet{bus10} the fractional
areas $f_u$ and $f_d$ occupied by the upward and downward mass
transport, respectively, have been set to be $f_u = f_d (\equiv f)$
= 0.5. When this analysis is pursued in the framework of a slow
mixing process, the physical mechanism of transport and its velocity
do not play any crucial role; the nucleosynthesis results are then
controlled by the path integral of reaction rates over the
trajectory. If needed, the mixing velocity $v$ can then be simply
derived as: $v = {1\over f} \times {{\dot M} \over {4 \pi r^2 \rho}}$. As
discussed elsewhere \citep{gua9}, very different scenarios may open
up when the transport is operated by fast mechanisms, suitable to
yield mixing time scales comparable to (or shorter than) the $^7$Be
half-life. The complex processes associated to the nucleosynthesis
of low mass nuclei like D, $^3$He, Li, Be and B will however be
postponed to another work.

After core He-exhaustion, our low-mass stellar models evolve to the
AGB  stages. General discussions of the physics of AGB stars can be
found in \citet{ir83,bgw,sgc}. On the AGB, deep mixing phenomena can
occur again, and there is actually clear observational evidence of
their presence in those final stages \citep{bus10}.  In fact, below
the convective envelopes of these evolved red giants, conditions of
chemical homogeneity similar to those found after FDU are
established. During the early phase of the AGB (before thermal
pulses start) this is guaranteed by the previous advancement of the
H-burning shell during the first ascent to the RGB. Subsequently,
when thermal instabilities from the He shell begin, the downward
envelope expansions (which follow them) easily reach down to the
H-He discontinuity, even before the penetration into the He-rich
layers occurs, a fact that formally defines the third dredge up, or
TDU. In most of the interpulse periods of AGB phases, therefore, H
burning proceeds in an initially-homogeneous environment, i.e. in
conditions suitable for the occurrence of extra-mixing.

\begin{figure*}
\centering{\includegraphics[width=0.5\textwidth]{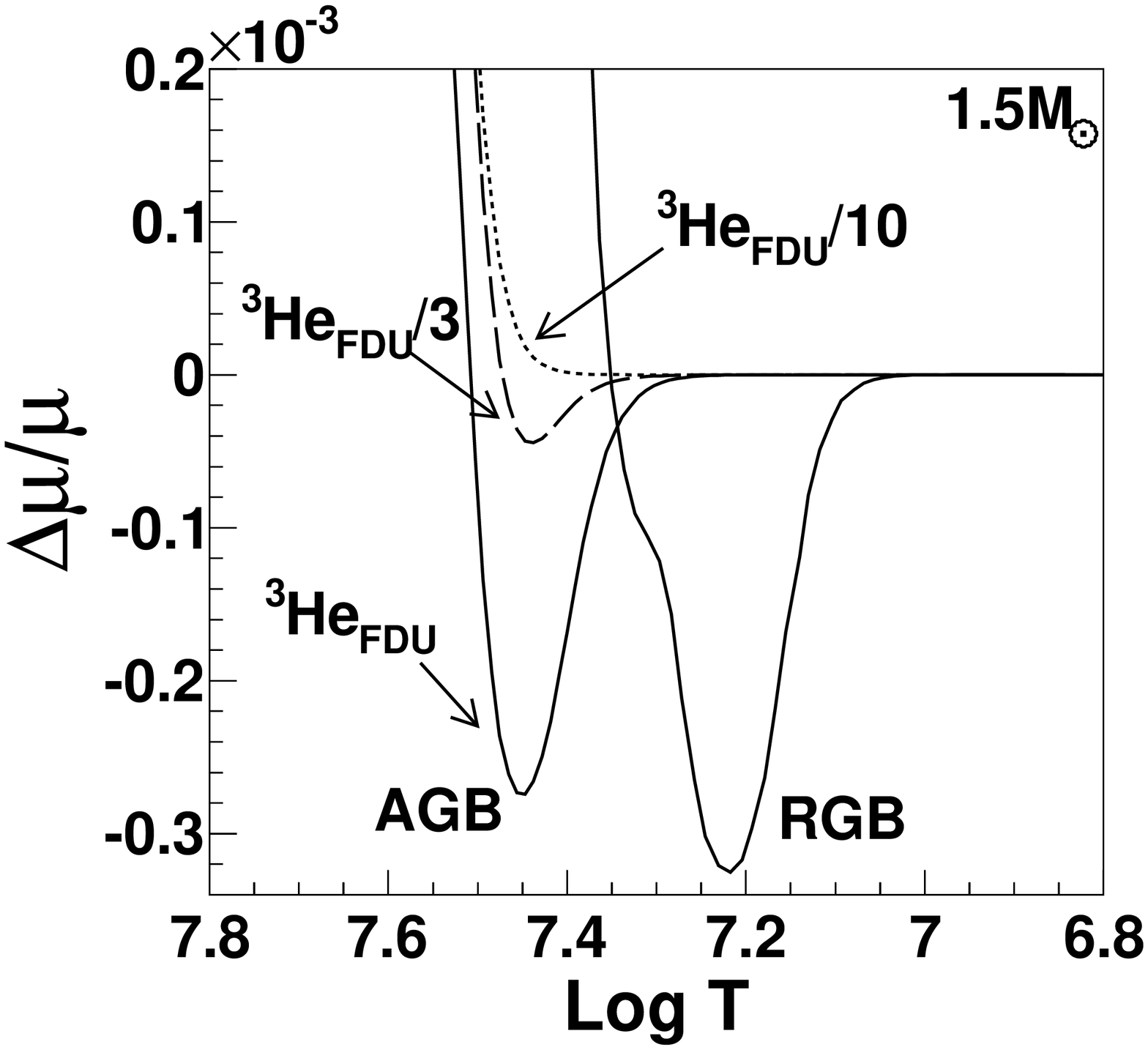}
\includegraphics[width=0.5\textwidth]{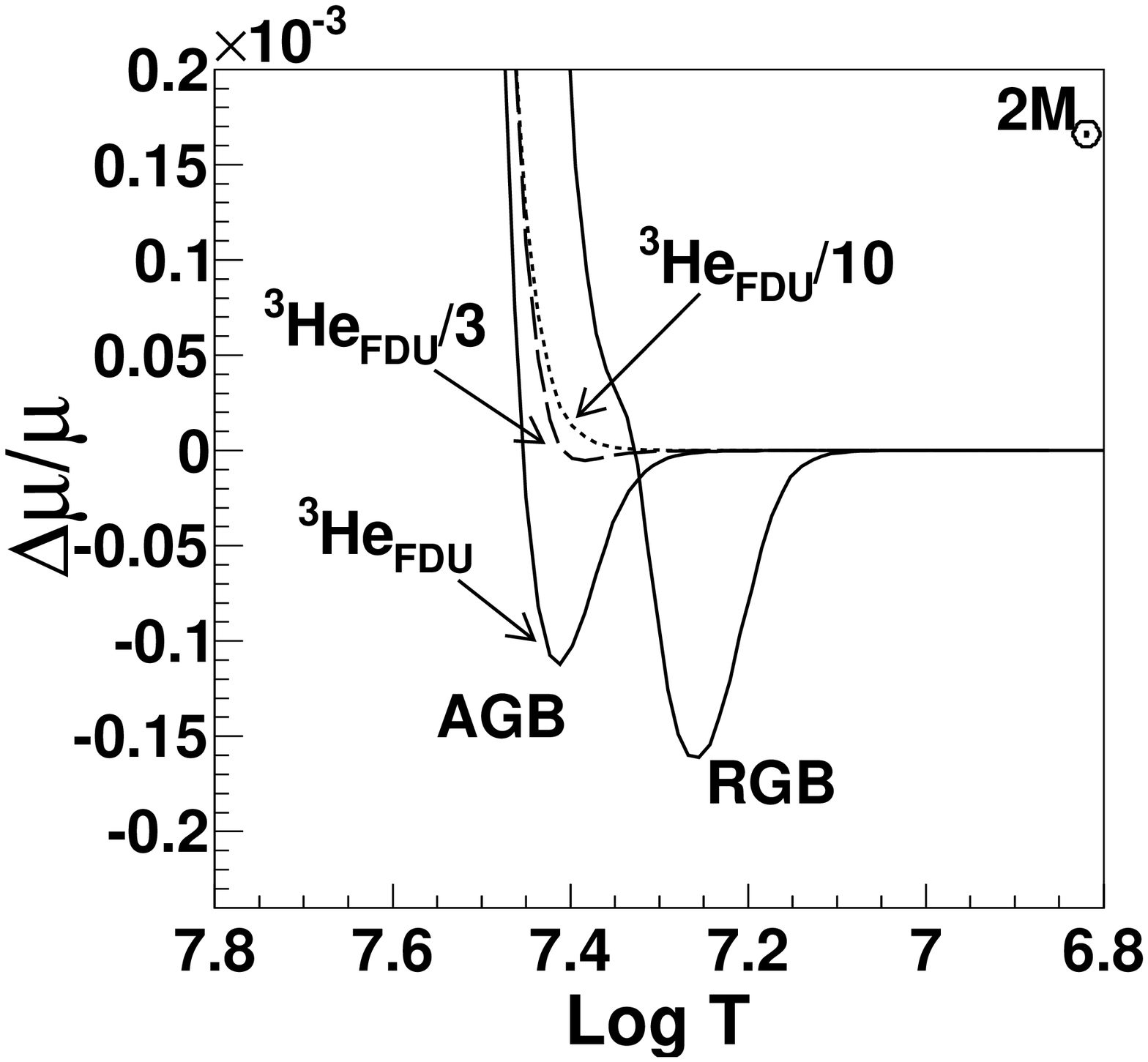}
} \caption{Relative variation of the molecular weight ($\Delta \mu /\mu$) for
a 1.5 and and a 2.0 M$_{\odot}$ star with solar metallicity, in the
layers hosting $^3$He burning above the H shell. The $\mu$
inversion, present on the RGB, is preserved also on the AGB only if
a sufficient supply of $^3$He remains. For the AGB, different lines refer to different
abundances of $^3$He resulting from the previous RGB phase, as indicated by the labels.} \label{f8}
\end{figure*}

Various experimental constraints (and in particular the spread in
the  C, O and Al isotopic ratios shown by presolar grains) suggest
that the details of the transport vary from star to star
\citep{nol03}. One has also to mention that not all the physical
mechanisms suitable to drive chemical transport in first-ascent red
giants are necessarily available also during the AGB stages. This
was discussed by \citet{bus10}, who showed that the operation of
deep mixing on the RGB, by whatever cause, consumes $^3$He, so that
the molecular weight inversion generated by its burning occurs in AGB phases only if the amount of $^3$He
remaining in the envelope is substantial \citep[as is the case in
the work by][]{cl10}. Otherwise, different physical mechanisms must
be looked for.
In Fig.\ref{f8} we better illustrate this point reporting the variation of the molecular weight ($\mu$)
across the radiative region of a 1.5 and and a 2.0 M$_{\odot}$
star with solar metallicity. While the $\mu$ inversion on the RGB is appreciable
in both the considered models, on the AGB it is preserved only if a sufficient supply of $^3$He remains.
In particular, for the AGB, different lines refer to different abundances of $^3$He resulting from the previous RGB
phase: no destruction, a consumption by factors of 3 or 10. It is evident that the stronger the previous
$^3$He consumption is, the weaker is the $\mu$ inversion.

\section{Results for the RGB phases}

\begin{table}
\centerline{Table 1}
\centerline{Nitrogen and Oxygen Isotopes at First Dredge Up.}
\centerline{(1.5 M$_{\odot}$ models with updated reaction rates)}
\centerline{
\begin{tabular}{cccccc}\hline
Case & $^{15}$N/$^{14}$N & $^{17}$O/$^{16}$O & $^{18}$O/$^{16}$O\\
\hline
$Z = $0.01              & 6.17$\cdot$10$^{-4}$ & 1.44$\cdot$10$^{-3}$  & 1.57$\cdot$10$^{-3}$\\
$Z = $0.014 (${\odot}$) & 6.27$\cdot$10$^{-4}$ & 1.21$\cdot$10$^{-3}$  & 1.59$\cdot$10$^{-3}$\\
$Z = $0.02              & 6.65$\cdot$10$^{-4}$ & 9.99$\cdot$10$^{-4}$  &  1.62$\cdot$10$^{-3}$\\
\citet{nol03} & 1.81$\cdot$10$^{-3}$ & 8.10$\cdot$10$^{-4}$ & 1.62$\cdot$10$^{-3}$\\
\hline
\end{tabular}\label{fdu}
}
\end{table}

Table 1 illustrates the main changes in the nitrogen and oxygen
isotopic ratios induced by FDU, for the sample case of a 1.5
M$_{\odot}$ star. In order to allow for comparisons with previous
results and to disentangle the reasons of the changes, the table is
presented for the solar metallicity and for the typical cases $Z =
0.01$ and $Z = 0.02$, often used in deep mixing discussions. As a
comparison, the values previously adopted by \citet{nol03}, taken
from \citet{bs99} for the same stellar mass are also shown.

\begin{table}
\centerline{Table 2}
\centerline{Nitrogen and Oxygen Isotopes at FDU}
\centerline{(Variation with reaction rates, $Z = Z_{\odot}$)}
\vskip 0.3cm
\centerline{a): updating only the $^{14}$N(p,$\gamma$)$^{15}$O}
\centerline{
\begin{tabular}{cccccc}
\hline
Mass (M$_{\odot}$)& $^{15}$N/$^{14}$N & $^{17}$O/$^{16}$O & $^{18}$O/$^{16}$O\\
\hline
 1.0 & 2.07$\cdot$10$^{-3}$ & 4.29$\cdot$10$^{-4}$ & 2.42$\cdot$10$^{-3}$ \\
 1.2 & 1.51$\cdot$10$^{-3}$ & 5.44$\cdot$10$^{-4}$ & 2.19$\cdot$10$^{-3}$ \\
 1.5 & 1.09$\cdot$10$^{-3}$ & 1.26$\cdot$10$^{-3}$ & 2.01$\cdot$10$^{-3}$ \\
 1.7 & 9.16$\cdot$10$^{-4}$ & 2.30$\cdot$10$^{-3}$ & 1.91$\cdot$10$^{-3}$ \\
 2.0 & 7.79$\cdot$10$^{-4}$ & 4.87$\cdot$10$^{-3}$ & 1.83$\cdot$10$^{-3}$ \\
\hline
\end{tabular}
}
\vskip 0.3cm
\centerline{b) ADE10 \& ILI10 rates; new [$^{15}$N/$^{14}$N]$_{\odot}$}
\centerline{
\begin{tabular}{cccccc}
\hline
Mass (M$_{\odot}$)& $^{15}$N/$^{14}$N & $^{17}$O/$^{16}$O & $^{18}$O/$^{16}$O\\
\hline
 1.0 & 1.19$\cdot$10$^{-3}$ & 3.81$\cdot$10$^{-4}$ & 1.89$\cdot$10$^{-3}$ \\
 1.2 & 8.64$\cdot$10$^{-4}$ & 4.82$\cdot$10$^{-4}$ & 1.73$\cdot$10$^{-3}$ \\
 1.5 & 6.27$\cdot$10$^{-4}$ & 1.11$\cdot$10$^{-3}$ & 1.58$\cdot$10$^{-3}$ \\
 1.7 & 5.25$\cdot$10$^{-4}$ & 2.04$\cdot$10$^{-3}$ & 1.51$\cdot$10$^{-3}$ \\
 2.0 & 4.46$\cdot$10$^{-4}$ & 4.32$\cdot$10$^{-3}$ & 1.44$\cdot$10$^{-3}$ \\
\hline
\end{tabular}\label{fdu1}
}
\end{table}

Table 1 reveals that the estimates for the surface isotopic ratios
of nitrogen  and oxygen at FDU can change by large amounts with
respect to previous findings. In particular, this is true for
$^{15}$N/$^{14}$N and for  $^{17}$O/$^{16}$O. In order to show the
selective relevance of the model choices in producing these results,
Table 2 shows the same nitrogen and oxygen ratios as obtained: a)
updating, with respect to \citet{nol03}, the physical model
parameters, the value of the solar metallicity and the
$^{14}$N(p,$\gamma$)$^{15}$O rate; and b) adding also all the other
new reaction rates by ADE10 and ILI10, as discussed in section 2,
and using the new (Jovian) value for the solar system nitrogen
isotopic ratio. This is done for the masses 1, 1.2, 1.5, 1.7, 2
M$_{\odot}$, at solar metallicity.

\begin{figure*}
\centering
{\includegraphics[width=0.8\textwidth]{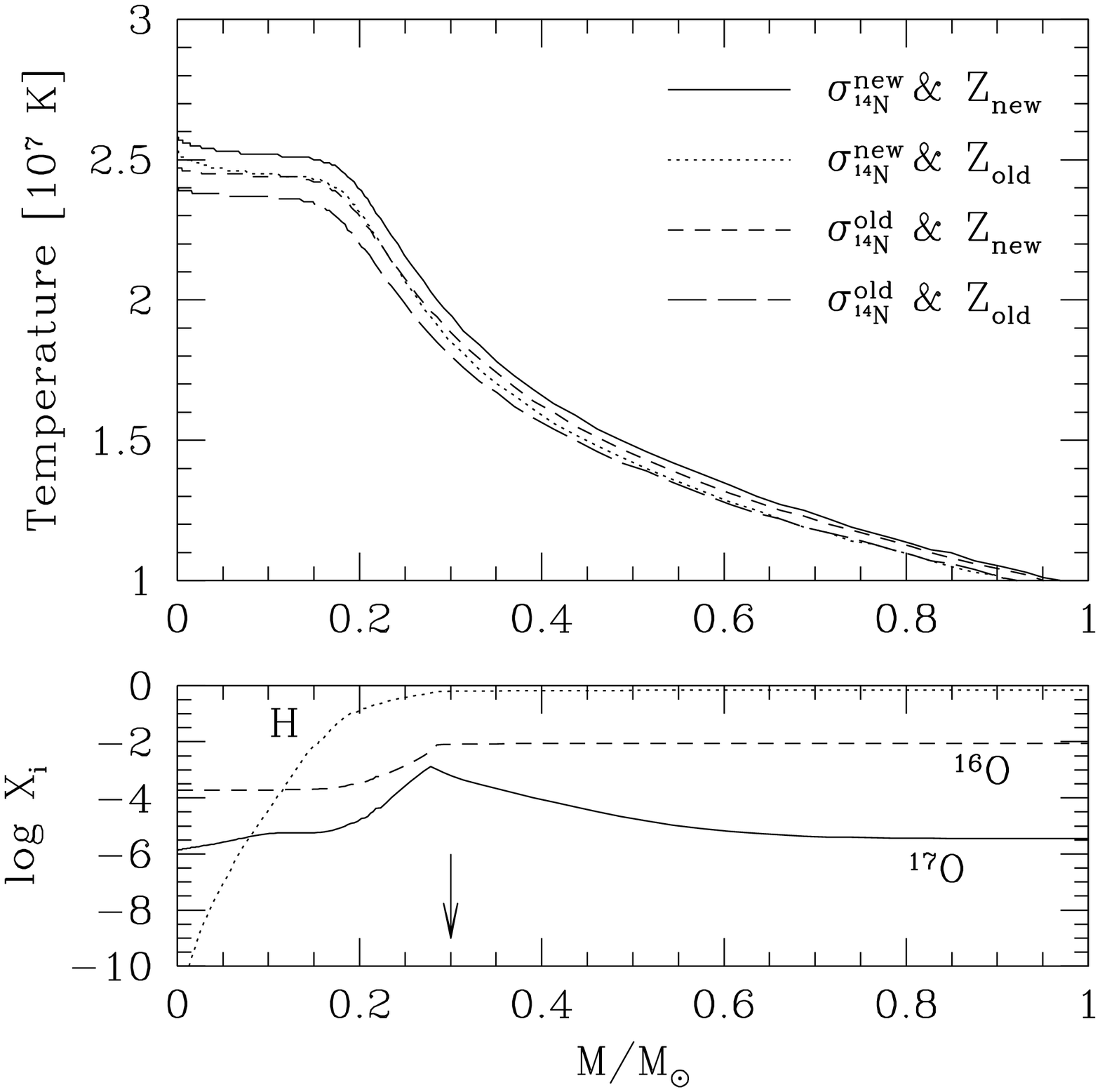}} \caption{Upper panel: the profile of
the stellar temperature as a function of the fractional mass at central H exhaustion in a model
star of 2 M$_{\odot}$, run with the indicated choices of
the reaction rate for proton captures on $^{14}$N and of the solar metallicity. $Z_{new}$ and $Z_{old}$ refer to the
values reported by \citet{as09} and \citet{ag89}, respectively.
Lower panel: the $^{16}$O and $^{17}$O abundance profiles at the same moment to which panel 1 refers.
The plot is for our standard case (i.e. has the new $^{14}$N(p,$\gamma$)$^{15}$O reaction rate and
the new solar metallicity). The arrow indicates the layer reached by FDU.} \label{f9}
\end{figure*}

A comparison between the two tables makes clear that, while for
nitrogen the change is mainly due to the choice of the new
solar value, the large shifts for $^{17}$O/$^{16}$O descend from both the new
abundances and the new rate for proton captures on $^{14}$N.
Indeed, both the mentioned upgrades induce small but effective changes in the stellar
structure, in particular in the temperature profile and in the maximum
penetration of the envelope during the FDU. These changes are obviously larger
for stars where CNO cycling is more efficient: our 2 M$_\odot$ model has $^{17}$O/$^{16}$O
= 5.14$\times 10^{-3}$, which is more than twice the value of 2$\times 10^{-3}$, previously
used in deep mixing models by \citet{nol03} and \citet{bs99}. In this case, the new adopted solar
metallicity increases the $^{17}$O/$^{16}$O by about 65\%, while the
new $^{14}$N(p,$\gamma$)$^{15}$O adds a further 35\%.
In the upper panel of Fig \ref{f9} we report the temperature profile at
central hydrogen exhaustion characterizing four different 2 M$_\odot$
models, run with different inputs:
\begin{enumerate}
\item{standard case (we shall call it ST); it has the new $\sigma_{^{14}\rm N}$ and
$Z=Z_{\odot}$;}
\item{test case with the new $\sigma_{^{14}\rm N}$ and $Z$=0.02 \citep[this is the solar metallicity derived by
adopting the compilation of][]{ag89};}
\item{test case with the old $\sigma_{^{14}\rm N}$ and $Z=Z_{\odot}$;}
\item{test case with the old $\sigma_{^{14}\rm N}$ and Z=0.02.}
 \end{enumerate}
The standard case shows the highest temperature profile, this fact
 leading to a larger $^{17}$O production in the region between 0.3 and 0.8
 M$_\odot$ (the inner core is not affected by the FDU, while the external layers
 have temperatures too low to activate H-burning). In the lower
 panel of Fig.\ref{f9} we report the abundances of hydrogen, $^{16}$O and $^{17}$O
of our ST case. In the same panel we also report the maximum
penetration of the convective envelope during FDU.
% (marked by the arrow).
 Note that, although the envelope penetrates less than in the other tests
 (M$_{bottom}^1$=0.302, M$_{bottom}^2$=0.290, M$_{bottom}^3$=0.298,
 M$_{bottom}^4$=0.286), the standard case shows the largest
 $^{17}$O/$^{16}$O value, because of the strong dependence of this ratio on the
 temperature \citep[see also][their Fig. 8 in particular]{nol03}.

A look at Table 2 reveals that the other reaction rate changes of section
2, including those directly affecting oxygen isotopes, induce
smaller (although non-negligible) effects than the
$^{14}$N(p,$\gamma$)$^{15}$O does, in altering the isotopic mix of
oxygen at FDU.

\begin{figure*}
\centering{\includegraphics[height=10cm]{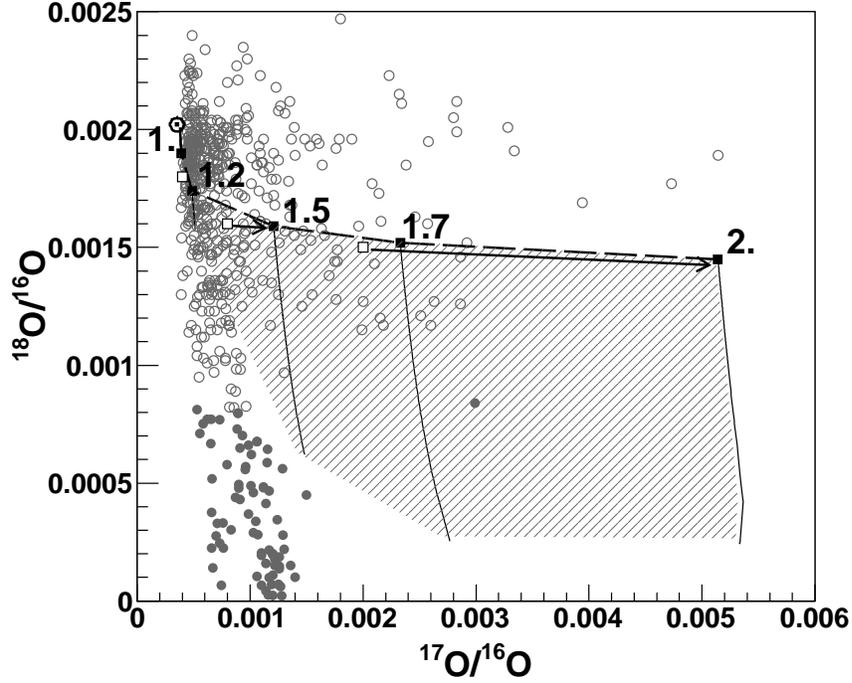}} \caption{The
oxygen isotopic mix in the envelope of our solar-metallicity stars
at FDU (solid line). Full squares indicate the mass of the model
star. As a comparison, open symbols (see arrows for associations)
show the composition for the masses 1.2, 1.5 and 2 M$_{\odot}$ in
the old scenario with NACRE rates and with the solar metallicity
from \citet{ag89}. Dashed lines (and the corresponding shaded area)
show the range of values covered by deep mixing on the RGB, for
$\Delta$ = 0.15 and $\dot M_6$ = 0.3. These are rather extreme
values of the parameters for the RGB phase, aimed at showing the
maximum effects. The grey data points refer to measurements in
presolar grains \citep[][WUSTL Presolar Database $http://presolar.wustl.edu/~pgd/$]
{choi,nit97,nit08}. We plot those of group 1
(open circles) and of group 2 (filled circles). The solar symbol
shows the initial isotopic mix. It is clear that extra-mixing on the
RGB, even operating with extreme efficiency, does not suffice to
explain the data of group 2 grains.}\label{f10}
\end{figure*}

Fig. \ref{f10} shows this in graphic form: filled squares represent
the isotopic composition of oxygen at FDU in our models,
while open squares show the previous situation, with the old NACRE
rates and the old choice for the solar metallicity. The new results
confine the stellar masses that produced presolar oxide grains
within a very narrow range ($1 \le M/M_{\odot} \le 1.5 - 1.7$).

The abundances seen at the surface of a giant star near the tip of
the RGB are  then known to differ from those at FDU for the effects
of deep mixing phenomena, which change the envelope composition by
linking its material to the hotter, radiative regions above the
H-burning shell, where proton captures re-establish a varying
abundance distribution, starting from an initially-homogeneous
composition. Fig. \ref{f10} also shows the effects of deep mixing
on the RGB (for a specific choice of the parameters defined in
Section 4, namely $\Delta$ = 0.15 and $\dot M_6$ = 0.3). As the
figure shows, presolar grains with a moderate $^{18}$O destruction,
mainly of group 1 \citep{nit97}, can be reproduced by this rather
extreme processing, provided the stellar masses are confined within
the range mentioned above. This range is the same where most MS-S
stars are found. We shall come back on this point in the next
section. Most grain data, however, and especially those with both
$^{18}$O/$^{16}$O and $^{17}$O/$^{16}$O ratios below 0.0012,
 require even more extended processing than
presented in the figure and clearly indicate that deep mixing cannot
occur only on the RGB \citep[see][]{bus10}.

\begin{figure*}
\centering
\includegraphics[width=0.8\textwidth]{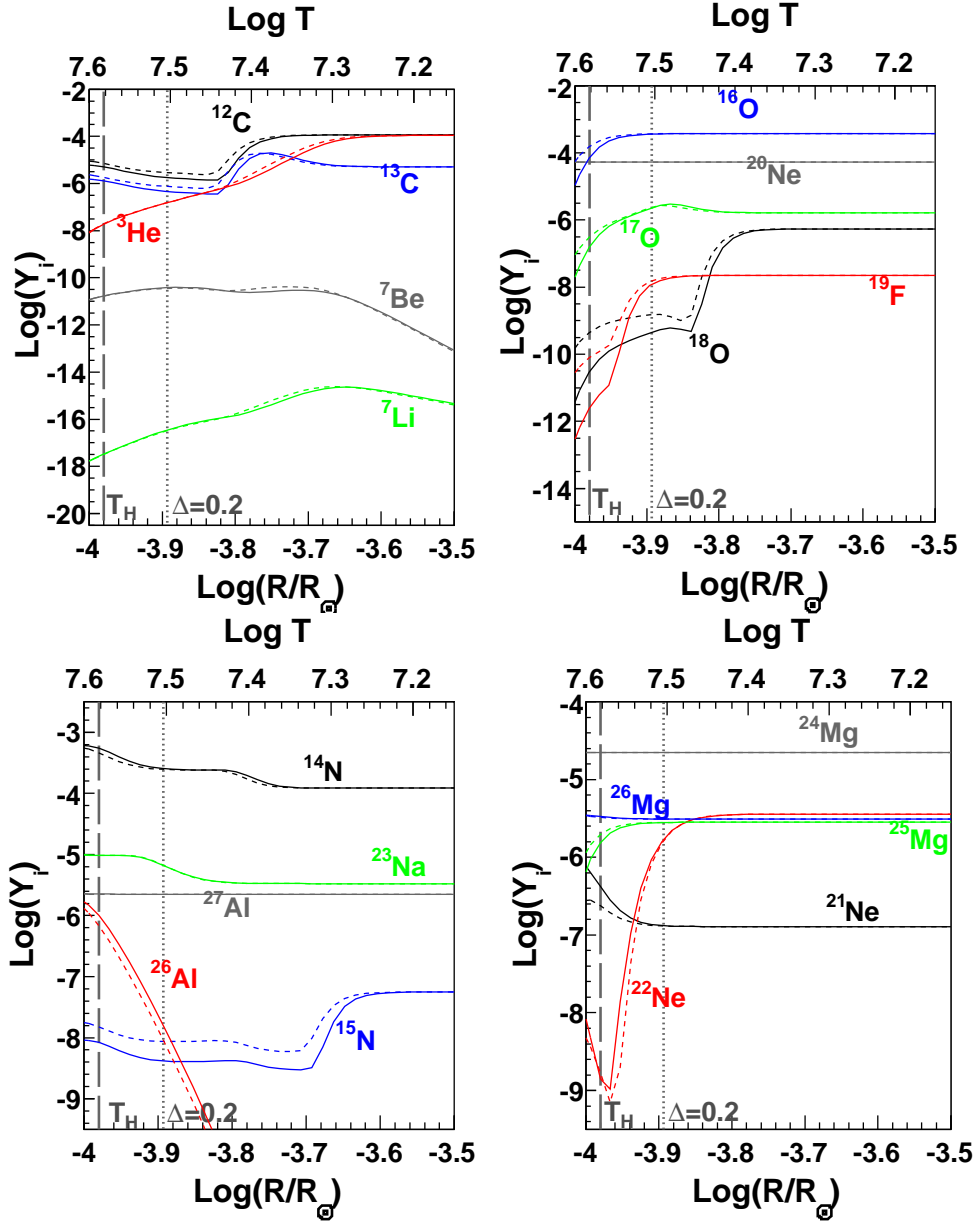} \caption{Distribution of abundances
for some light and intermediate-mass isotopes  in the radiative
layers between the H-burning shell and the convective envelope. The
plots are taken shortly after the moment when shell-H burning
reaches the regions homogenized by FDU and re-establishes chemical
gradients there. Dotted lines in the figure
represent the old results, obtained before our reaction rate
updates, while continuous lines represent our new estimates. The
vertical dotted line and the dashed one indicate, respectively, the place where $\Delta$ = 0.2
and where the energy from the H-burning shell is maximum (T$_{\rm H}$).
 (A color version of this figure is available in the online journal.)}\label{f11}
\end{figure*}

Typical abundance profiles created by nucleosynthesis in the region
below the  convective envelope after the luminosity bump are shown
in the four panels of Fig. \ref{f11} taken from our 2 M$_{\odot}$ model
with initially solar composition. In regions below $\Delta \simeq 0.2$ (corresponding to
the vertical dashed line) the $^3$He+$^4$He reaction starts to dominate over the
$^3$He+$^3$He one (that induces the inversion of the molecular weight);
this is therefore the innermost position from which thermohaline diffusion can start. Notice that some
remarkable abundance changes, including the production of a critical
nucleus like $^{23}$Na, and an enhanced destruction of $^{18}$O,
occur immediately below this zone. If only thermohaline mixing were to take
care of transporting matter to the envelope on the RGB, then we would
confirm the recent findings by \citet{cl10}, according to which
these nuclei are only moderately affected in RGB stages. If instead
deeper mixing mechanisms were at play, this result would change
drastically (as we have shown for the destruction of $^{18}$O in
Fig. \ref{f10}).

\begin{figure*}
{\centering {\includegraphics[height=6.5cm]{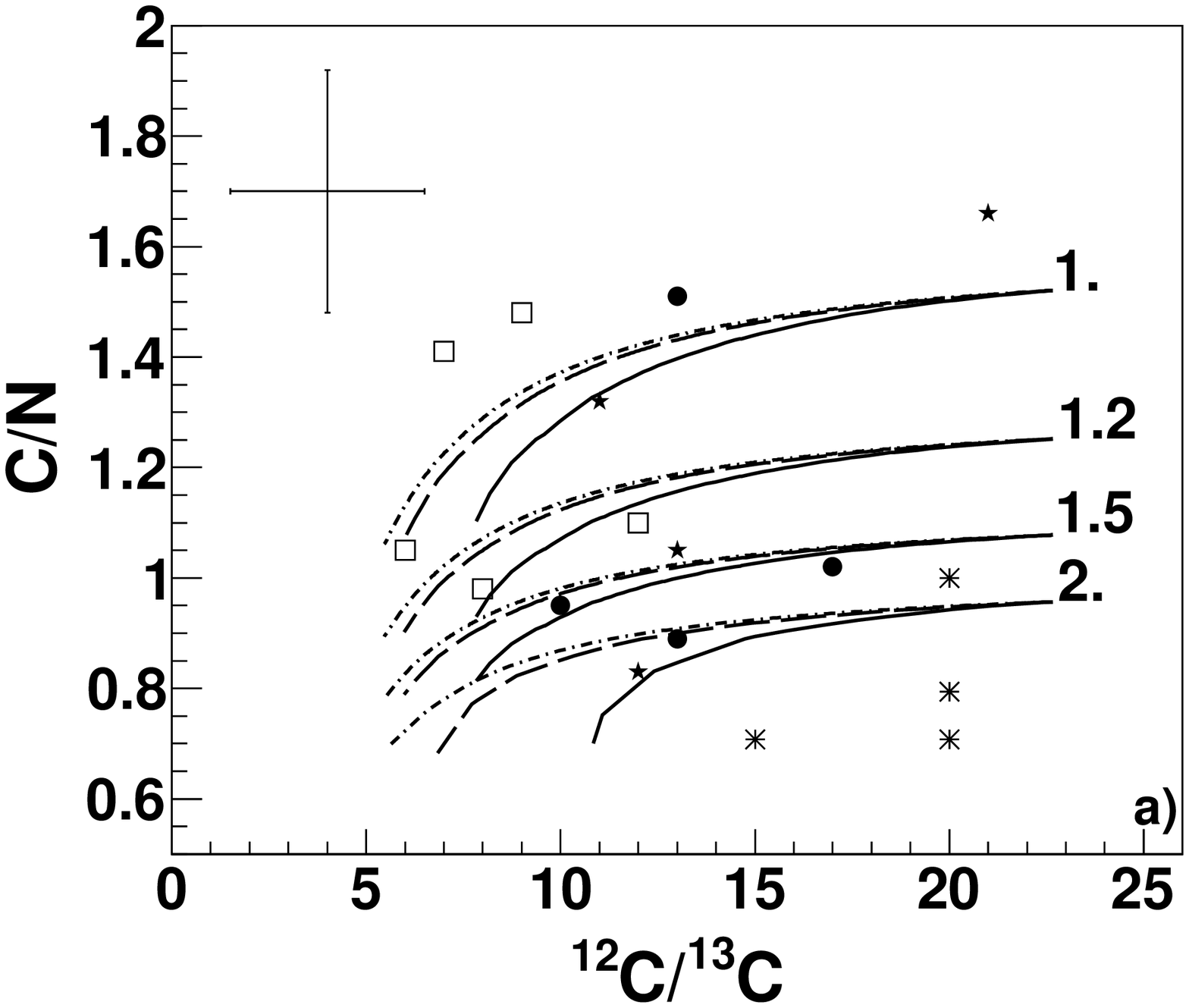}
\includegraphics[height=6.5cm]{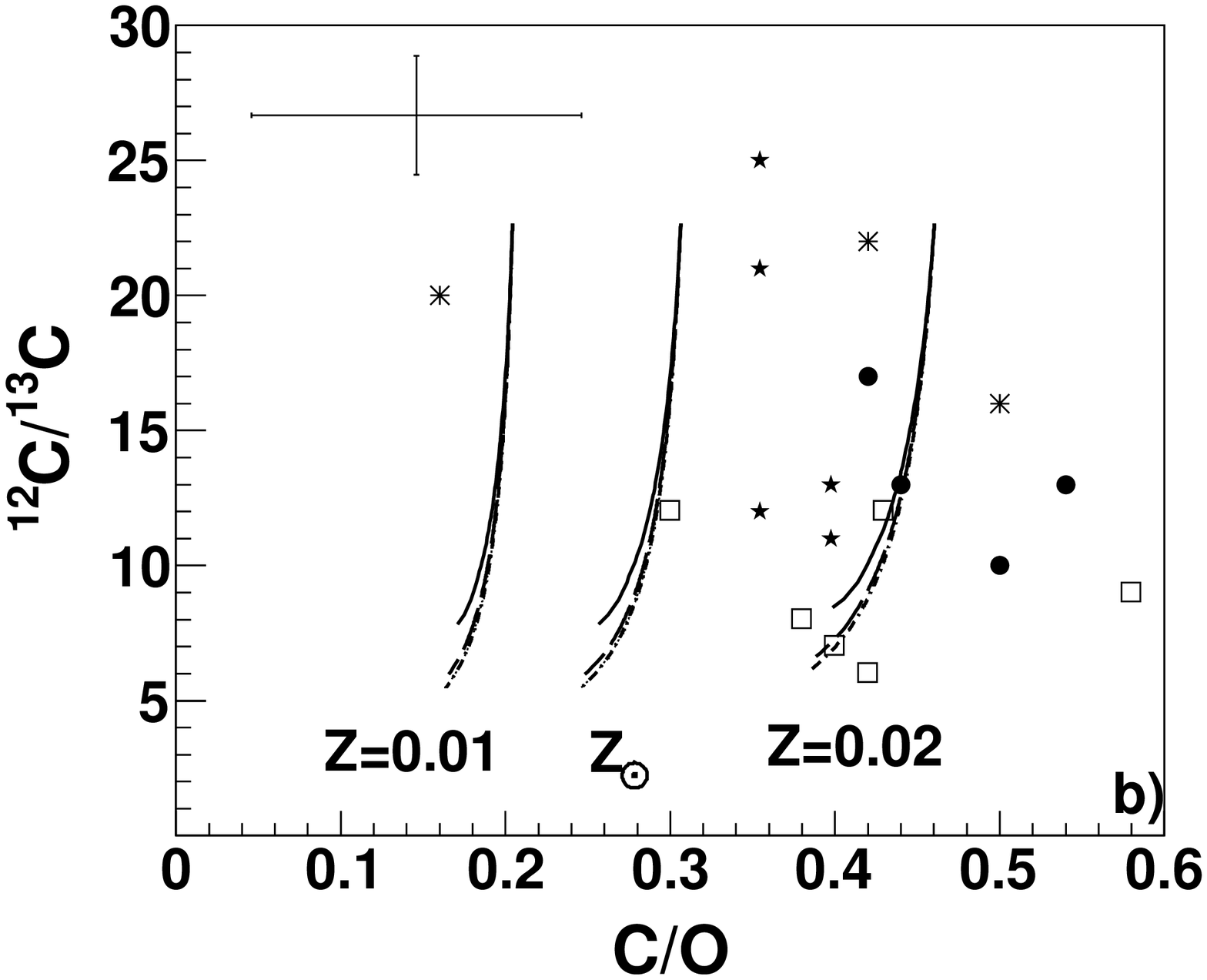}
\includegraphics[height=6.5cm]{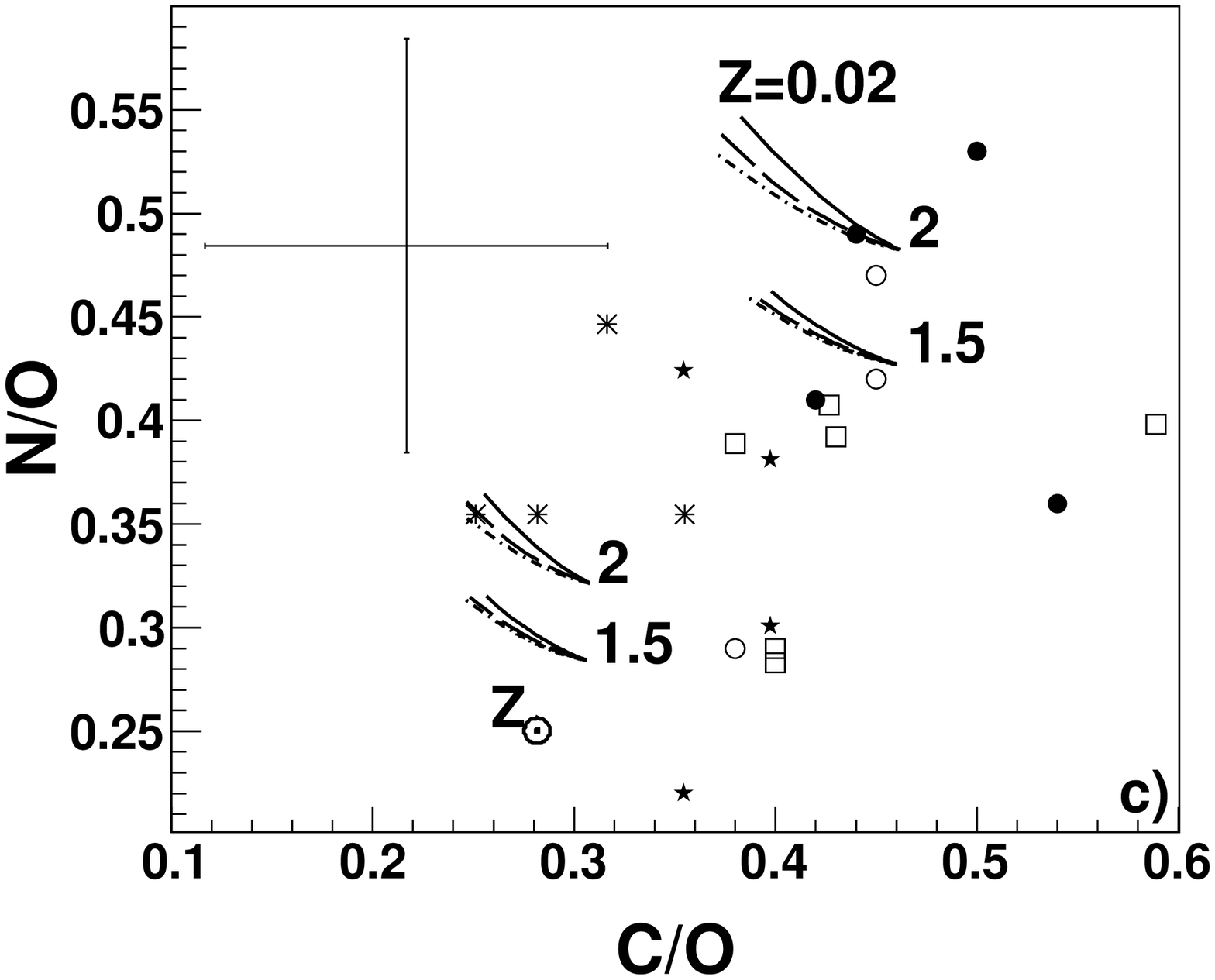}}}
\caption{Deep-mixing results for CNO nuclei, as compared to
observations in RGB  stars of metallicities close to solar. Panel
a): model curves correspond to $\Delta$ = 0.2; they proceed from
right to left, starting from abundances at FDU. Continuous lines
refer to $\dot M_6$ = 0.3, dotted lines to $\dot M_6$ = 0.1, dashed
lines to $\dot M_6$ = 0.03. Black stars show post-FDU stars from M67
\citep{b87,gb91}; asterisks are data from \citet{cot}; open squares
refer to open clusters by \citet{mik10}. Circles refer to data from
\citet{smil9}, in particular to NGC5822 (filled symbols) and NGC2360
(open symbols). Only solar metallicity models are shown, to avoid
excessive crowding; the cases at $Z = 0.02$ would be slightly
shifted upward. The typical uncertainty (1$\sigma$) is indicated.
Panel b): same type of comparison, this time for the carbon isotopic
ratio versus C/O. Models of $Z = 0.02$ and $Z= 0.01$ have been
included. Panel c): same comparison, for the N/O versus C/O
ratios.}
\label{f12}
\end{figure*}

Indication on the mixing depth can obviously be derived by
observational data.  We must however recall that spectroscopic
measurements, especially for isotopes, have very large
uncertainties, so that their constraints should be used with care.
Furthermore, the metallicity of Population I  red giants spans over
a large range, while our models cover only a factor-of-two interval.
As a consequence, among the several existing compilations of CNO
abundances for red giants of the galactic disk we can consider only
those with metallicities in the range 0.01 $\le Z \le 0.02$, as the
interplay of nucleosynthesis and mixing is extremely sensitive to
metallicity variations. The best comparisons for our
solar-metallicity cases ($Z = 0.014$) should come from M67, whose
composition has always been confirmed to be very close to solar
\citep{ran06}. For this cluster CN abundances and $^{12}$C/$^{13}$C
ratios were determined by \citet{b87} and \citet{gb91}, but the
oxygen abundance was not measured and rather assumed from older
works. New determinations of the oxygen concentration in M67 now
exist \citep{t00,y05,p10} and their [O/Fe] ratios essentially agree
with one another; however, other rather different suggestions are
also available \citep{fr10}. We shall adopt here the [O/Fe] and
[Fe/H] values by \citet{y05}, rescaling oxygen for our updated
choice of the solar abundance from \citet{as09}. The results of our
$Z = 0.02$ case can instead be compared to observations in evolved
RGB stars of open clusters younger than the Sun, like e.g. those
studied by \citet{smil9} and \citet{mik10}. Only few  stars from samples of the
old disk \citep[see e.g.][]{cot,shet} can instead be used for our
purposes, namely those whose metallicity is close to solar.
Fig. \ref{f12} shows a comparison of our models with observations selected
according to the above criteria.  Given the uncertainty of
spectroscopic measurements (especially the older ones), the
agreement is quite good, especially for M67. The observations do
require deep mixing at a moderate temperature ($\Delta = 0.2$) and
at circulation rates 0.1 $\le \dot{M_6} \le$ 0.3.
From the value of $\Delta$ we infer that
thermohaline diffusion might be adequate for this purpose. However,
warnings on the mixing velocity and on the diffusion coefficient
deduced from recent literature have to be considered and we refer to
section 7 for a more detailed discussion of that.

\section{Results for the AGB phases}
For the late evolutionary stages, the adoption of low temperature
C-enhanced opacity  tables, the choice of a new mass loss law
(limiting the number of thermal pulses) and the different treatment
of the inner convective border all play important roles in
controlling the final results. While, again, we direct the reader
to \citet{cri09a} for details, we need to recall  here that the
present AGB models have now a much
stronger support from observations. In particular, the new opacities
and the more effective mass loss law adopted combine in inducing
lower values for the surface temperature and luminosity. In their
turn, these facts affect the colors, the infrared emission and the
luminosity function of AGB stars. The new models now succeed in
reproducing the photometric and spectroscopic properties of their
observational counterparts; in particular, the low effective
temperatures of C stars at different metallicities
\citep{dom04,bus07b} can now be naturally reproduced. The stronger
mass loss rates reduce the number of thermal pulses, but the
introduction of an exponentially decaying profile of convective
velocities at the base of the envelope increases the TDU efficiency,
so that the final enhancement factor of the most significant nuclei
produced by AGB stars is roughly the same as before; but now this is
obtained together with a good reproduction of the luminosity
functions of AGB stars \citep[][in preparation]{cri11}.

We also recall that the adopted algorithm for the velocity profile
accounts,  after each TDU episode, for the formation of a tiny
region enriched in protons where, subsequently, $^{13}$C can be
formed. This is the radiative $^{13}$C pocket where ($\alpha$,n) captures
on $^{13}$C provide the main neutron source for the synthesis of heavy n-rich nuclei in
AGB stars \citep[see e.g.][]{bgw,lu03,sgc,sne08}, thus complementing the production of lighter
n-rich nuclei in massive stars \citep[see e.g.][]{bg5,ra93,ka94,pi10}. We however postpone the
analysis of this important point to a future paper.

\begin{figure*}
\centering{\includegraphics[width=0.8\textwidth]{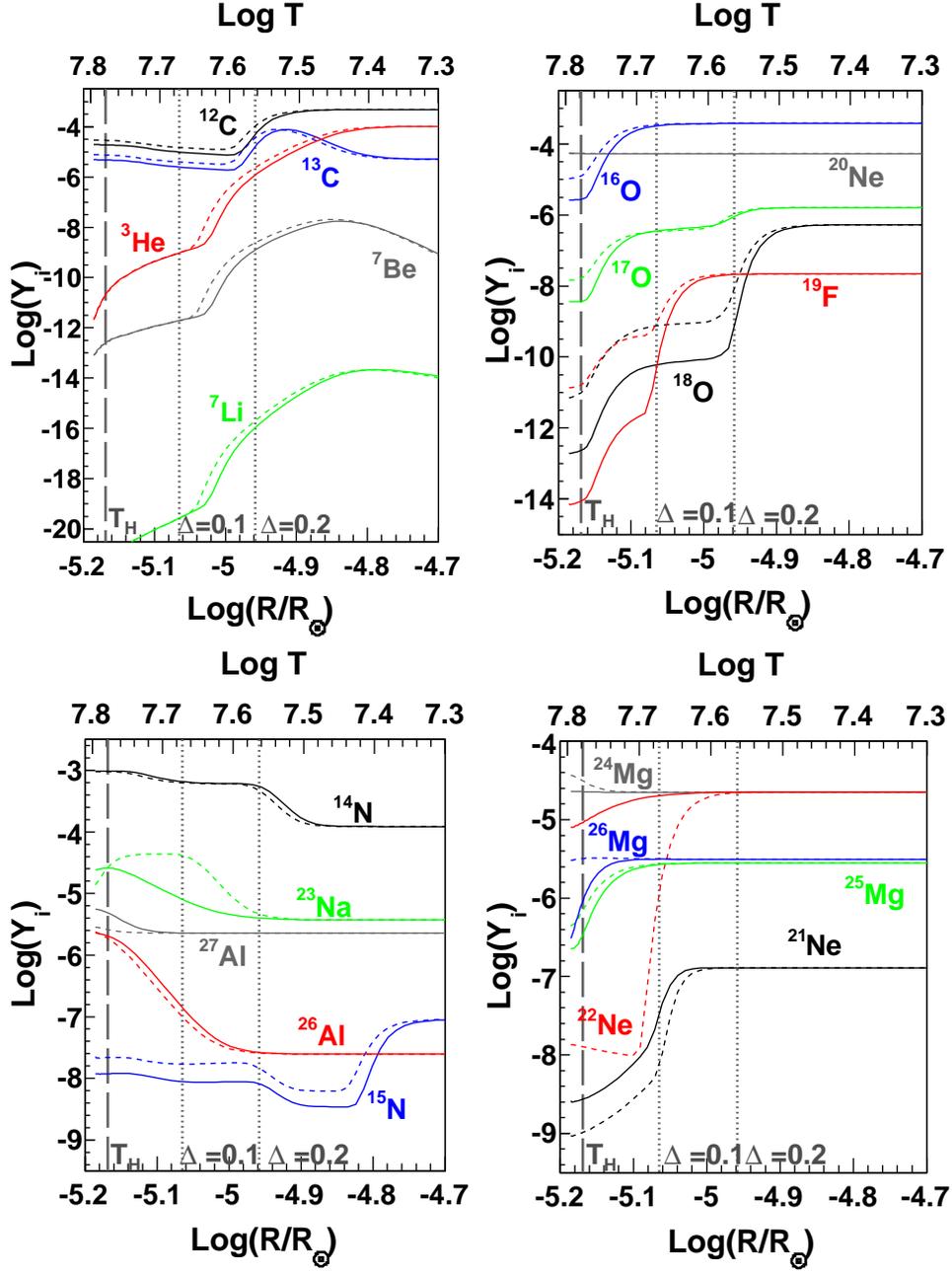}}
\caption{Distribution of abundances for
some light and intermediate-mass isotopes in the radiative layers
between the H-burning shell and the convective envelope. The plots
are taken roughly at mid-AGB evolution. The
two vertical dotted lines and the dashed one indicate the places where $\Delta$ = 0.2, 0.1 and 0.0, respectively.
This last position corresponds to the layer where the energy from the H-burning shell is maximum (and $T$ = $T_{\rm H}$).
 (A color version of this figure is available in the online journal.)} \label{f13}
\end{figure*}

Deep mixing in thermally-pulsing AGB stages operates on the
composition established above the H-burning shell and homogenized
after each pulse by the penetration of the convective envelope.
Typical abundances established in those layers by proton captures
are shown in Fig.  \ref{f13}. Previous work \citep{nol03,bus10} leads us
to expect that extra mixing mainly operates down to regions where
$\Delta$ = 0.1 to 0.2. These zones can be identified in Fig. \ref{f13}
from the two vertical lines. As is clear from the plots, remarkable
nucleosynthesis changes are induced there by the adoption of the new
reaction rates (continuous curves). Among the differences, we find a
larger destruction of $^{3}$He, $^{15}$N, $^{18}$O, $^{19}$F and a
lower production of $^{23}$Na. This last result is directly linked to
a lower destruction of $^{21}$Ne and $^{22}$Ne. The important
radioactive nucleus $^{26}$Al is more effectively produced than
before, but only marginally. In general, these results can be
interpreted as due to the slight increase in the temperature induced
by the new (less efficient) rate for $^{14}$N(p,$\gamma$)$^{15}$O,
superimposed to the specific effects of the rate changes for heavier
nuclei, which have a more local and limited effect.

As a consequence of the model changes and of the reaction rate
updates, the combined operation of mixing and  nucleosynthesis leads
now to abundance changes in the envelope like those summarized in
Fig. 7 (for the 1.5 M$_{\odot}$ model) and in Fig. 8 (for the
2 M$_{\odot}$ model). For these examples we adopted, on the RGB,
$\Delta$ = 0.2, $\dot M_6$ = 0.3; on the AGB, $\Delta$ = 0.1, $\dot
M_6$ = 1. These choices only serve for the purpose of
illustration. A wider sample of cases for the AGB is presented in
the subsequent Fig. 9 and Fig. 10\footnote{For researchers interested to
comparisons with specific observations, looking for an even wider
series of cases, we can provide the nucleosynthesis results upon
request.}.

The above figures referring to the AGB clearly show a saw-tooth
trend in the  abundances of several nuclei. This is induced by the
sudden mixing with He-rich layers at the third dredge up (TDU),
after each pulse. This trend is therefore pronounced for those
nuclei that are largely affected by He-shell nucleosynthesis, in
particular $^{12}$C, $^{22}$Ne and $^{26}$Al.

The $^{12}$C-enrichment by TDU is partly compensated by its
destruction in deep-mixing processing during the inter-pulse stages.
The operation of such mixing at even moderate efficiencies has the
consequence of preventing the formation of a carbon star in our
models up to 1.7 M$_{\odot}$. Combining this with the changes
already discussed for the FDU (see Fig. \ref{f10}), we can say that both
MS, S giants and oxide presolar grains should be produced mainly by
stars below this mass limit. Obviously, a more massive star (2
M$_{\odot}$  or more), before becoming C-rich, also passes through
an O-rich phase  in which TDU induces MS-S spectral characteristics;
but the high efficiency of dredge-up we have now implies that this
is a not a dominant phase. This can be inferred from the limited amount of mass lost
in the O-rich stages by a 2 M$_{\odot}$ model star becoming C-rich. For moderate
extra-mixing ($\dot M_6$ = 0.1)
this amounts to 0.4 and 0.2 M$_{\odot}$ at solar metallicity and at $Z=0.01$, respectively.
By contrast, more than  1 M$_{\odot}$ is lost
in the C-rich phase. On the contrary, a 1.5 $-$ 1.7 M$_{\odot}$ star looses its whole
envelope (at least 0.8 M$_{\odot}$) in the form of O-rich material. If one also considers
that the initial mass function favors these lower masses, one can deduce that
most of the observed MS and S giants should have a mass smaller than for C-stars and that
among these low mass AGB stars one should find the typical progenitors of presolar oxide grains.
There is obviously a possibility of finding MS-S giants among intermediate mass stars
($M \ge 4 M_{\odot}$), where hot bottom burning prevents the formation of
a carbon star. However, for them the expected isotopic ratios (especially for Oxygen) are very peculiar (e.g. with large excesses of $^{17}$O). \citet{ili08}, by comparisons of intermediate-mass models with presolar grain abundances, convincingly excluded any
connection of IM stars with presolar grains; presently, we don't know of even one single grain really requiring an intermediate-mass progenitor.

We can see this in some more detail. As already stated in discussing the RGB phases and in \citet{bus10}, oxygen isotopes in presolar oxide grains provide the most stringent constraints to stellar models with deep mixing, due to the fragility of $^{18}$O and to the sensitivity of $^{17}$O to temperature. Fig. \ref{f18} (left panel) presents these
constraints, showing the oxygen isotopic ratios in group 1 and group 2
grains together with model curves. These last include typical trends
for extra-mixing on the RGB at a moderate efficiency, plus the
results of deep-mixing computations for the AGB, for various choices
of the parameters. Notice that, by including the 1 M$_{\odot}$ star, the whole range of
$^{17}$O/$^{16}$O ratios is covered. This is another effect of the
model changes and corrects previous findings, based on older models,
according to which the grains at the extreme left would not be
explained by deep mixing at solar metallicity \citep{w5,nol03}. The
main reason of the change lays in the more marked destruction of
$^{18}$O at moderate temperatures, where $^{17}$O is still untouched
(see Fig. \ref{f11}, second panel). This is found also in very low-mass
models and allows their deep-mixing curves to proceed almost
vertically, covering a previously unaccounted area (not only the new
rate for the $^{14}$N(p,$\gamma$)$^{15}$O, but also those for
proton captures on $^{17}$O are important in determining this result).
The left panel of Fig. \ref{f18} also shows that the whole set of abundances for
$^{18}$O-poor grains (with $^{18}$O/$^{16}$O ratios below 0.0015) is explained by models for stars below about
1.7 M$_{\odot}$, those that will never become C rich.

The right panel of Fig. \ref{f18} then presents the  $^{17}$O/$^{16}$O
ratio as a function of $^{26}$Al/$^{27}$Al, for those grains that
show $^{26}$Mg excesses from $^{26}$Al decay. Here $\dot M_6$ is
assumed to be 3 (a rather efficient case).  Again,
most of the experimental results can be accounted for. A small
group of grains, for $^{17}$O/$^{16}$O ratios below 0.001 and
$^{26}$Al/$^{27}$Al above 0.01, are however not explained. We notice here
that no modification to the model might cover this area. In fact
(as an example) mixing at higher temperature would improve the fit to $^{26}$Al
but would worsen the problem for $^{17}$O (not to mention the fact that mixing at higher temperatures would affect the stellar structure). There appears to be a unique way out: an increase in the reaction rate for the $^{25}$Mg(p,$\gamma$)$^{26}$Al reaction at  temperatures sufficiently small to minimize the effects on $^{17}$O.
%In the energy region of our interest the main contribution to that reaction rate is due to the resonance at 93keV; we found %that,
%by artificially increasing its width by a factor of 3 to 5 the grains rich in $^{26}$Al and not in $^{17}$O can be accounted for (e.g. using $\Delta=0.05$ and %$\dot{M_6}=3$ in a 1.2 M$_{\odot}$ stellar model). We leave this suggestion about the resonance for future verifications.
%(Recent new measurements from LUNA are still under analysis, so that we should not wait a long time before knowing if this prediction can be confirmed).
%%%%
In the energy region of our interest the main contribution to that
reaction rate is due to resonances (at 58 and 93keV) for which large uncertainties are still present \citep{ili96}. We found that, by artificially increasing their strengths by a factor of 5 with respect to present recommendations,
%We found that, by artificially increasing this reaction rate by a factor of 5, in the energy region of our interest,
grains rich in $^{26}$Al and not in $^{17}$O can be accounted for (e.g. using $\Delta=0.05$ and $\dot{M_6}=3$ in a 1.2 M$_{\odot}$ stellar model). We leave this suggestion for future verifications. Recent new measurements (especially from LUNA) are still under analysis, so that we should not wait a long time before knowing if this prediction can be confirmed.

It is important to notice that the same presence of $^{26}$Al in the
grains, at a level higher than allowed by TDU (which therefore
requires the operation of deep mixing), poses crucial constraints to
the physical mechanisms driving the transport of processed matter.
Contrary to what was found on the RGB, here thermohaline diffusion
cannot be an explanation. Not only its effectiveness is put in doubt
(or at least reduced) by the previous $^3$He consumption, but the
layers where it starts are too cool to host any $^{26}$Al. This is
evident in Fig. \ref{f13}. On the AGB, the $^3$He+$^3$He reaction
prevails over the $^3$He+$^4$He for $\Delta \ge 0.15$, identified in
the first panel of the figure by the kink in the $^3$He curve. As
the third panel shows, the whole abundance of $^{26}$Al is produced
inside that zone. As we showed in Fig. \ref{f18}, accounting for
$^{26}$Al in oxide grains requires $\Delta \le 0.1$. We believe this
is  a clear demonstration that thermohaline mixing is not the
mechanism driving extra-mixing in AGB stars. Even its action as a
driver of magnetic-like mechanisms \citep{d09} would not be useful
in this case. Completely different physical processes, like e.g.
magnetic buoyancy \citep{bus10} or gravitational waves \citep{dt},
must necessarily be at play.

Further comparisons with observations are provided by Fig. \ref{f19}. Its
left panel shows  C/O and $^{12}$C/$^{13}$C measurements  in both
O-rich \citep{sl90} and C-rich stars \citep{abia02,abia10}. The curves refer to our 1.5
and 2 M$_{\odot}$ models; the more massive case is the one that
reaches C/O $\ge$ 1; its C-rich phases are marked by a continuous
line, the O-rich ones by a dashed line. We adopted $\Delta = 0.2$ and 0.25
and $\dot M_6$ values as indicated by the labels. It is clear that the
whole area of the observations is very well accounted for. This
figure offers a rather global test from stars, complementing those
derived from oxide grains; it confirms that the general scenario is
correct and that MS, S and C stars have indeed the masses we
suggested and do experience extra-mixing.

A much less satisfactory comparison is presented in the right panel
of Fig. \ref{f19}, where our model  curves are superimposed to the the
nitrogen and carbon isotopic ratios measured in SiC grains of AGB
origin \citep[see e.g.][and references therein]{zin1,zin2}. The
models in the figure are labeled by the values of $\Delta$ and
$\dot M_6$. They mainly cover the area of the so-called "mainstream"
SiC grains (with carbon isotopic ratios between 20 and 100), but only
for the upper part of the range in $^{14}$N/$^{15}$N. Models with
rather extreme $\dot M_6$ values can reach the area where the grains
with the lowest $^{12}$C/$^{13}$C ratios (called A \& B grains) are
present, but again only for very high values of $^{14}$N/$^{15}$N.
These inconsistencies are not new and have characterized the
attempts at explaining the nitrogen isotopes in SiC grains for many
years \citep[see e.g.][]{nol03}. Notice that the two filled circles,
at $^{14}$N/$^{15}$N = 680 and 5300, refer to the only (very
uncertain) data from red giants \citep{wan} and confirm that the RGB
ratios must indeed be much larger than solar. The low values
of many SiC grains have no explanation at all in low mass stars.
In principle, in the so-called $^{13}$C-pocket of AGB stars a reservoir of $^{15}$N
exists. This is generated trough proton and neutron captures on $^{14}$N and later, at the onset of the thermal pulse, by the activation of the $^{18}$O(p,$\alpha$) reaction, see also \citet{abia10}. However, these zones are not expected to be ejected and their $^{15}$N is subsequently destroyed in the convective thermal pulse.

Contamination of the grains with solar-system nitrogen seems unlikely, as there is no correlation between the $^{15}$N concentration and grain size, as would be expected (Zinner, private communication). A possible alternative is the contamination
from cosmic ray spallation, where $^{15}$N is produced abundantly.
In such a case the $^{15}$N content would be a measure of the grain
age and would not inform on stellar nucleosynthesis.

\section{Toward a physical model of deep mixing: constraints on the parameters}

As a concluding remark we want to make some comments on
the recent numerical simulations of extra-mixing presented by \citet{den10a} and by \citet{den10b}.
The work was performed through an analogy with current
research on thermohaline processes known to occur in the oceans.
The models were then extrapolated to the stellar case, on the basis
of the values of physical parameters valid for red giants; this treatment
has obviously a much higher degree of sophistication than in simple 1D analysis,
but still includes very uncertain parameters. Anyhow, the authors suggest
that the molecular weight inversion generated by $^3$He burning induces finger-like mixing
structures whose average aspect ratio (length over diameter, $\alpha = l/a$) is too small
by a factor of at least 50 as compared to what observations require; this aspect ratio,
squared, is proportional to the diffusion coefficient in a diffusive treatment, or to
the mixing velocity in a circulation-like transport \citep{nol03}. In \citet{den10a} it is
also mentioned that, for explaining the small number of red giant stars that are very Li-rich
($\log \epsilon$(Li) $\ge 1.5$) the aspect ratios found with thermohaline processes would
be too small by a factor of several hundreds; in fact, producing $^7$Li is known to require
a kind of transport fast enough to save to the envelope the short-lived $^7$Be ($t_{1/2}$ = 53 days).
These results would indicate that thermohaline diffusion alone is insufficient to
drive the chemical mixing required in red giants; as a way out, in \citet{den10b} it is suggested
that a coupling with magnetically-induced effects might be more promising.

The above results should now be verified by other authors, before final conclusions can be drawn.
Recent works either attribute the isotopic mix of red giants to the effects of thermohaline mixing alone \citep{cz} or
confirm that models for this process find velocities that are too low by a large factor, thus opening the possibility
that other mechanisms contribute \citep{cal10}.

Although our simple parameterized approach cannot enter into such details, we found the
treatment by \citet{den10a} and \citet{den10b} very instructive as a tool
to infer the required properties of suitable physical models for extra-mixing. In particular,
it was suggested \citep{bus07a,d09} that magnetic buoyancy might offer an alternative to
thermohaline diffusion, with the advantage that it would not require a nuclear process as
a trigger, hence it would have a wider possibility of application (e.g. to the formation of
the $^{13}$C-pocket). If magnetic buoyancy were to proceed through the development of $\Omega$-shaped
loops, formed as instabilities in MHD waves \citep{p94}, then each such structure would look like
a couple of mixing fingers. Let us make two illustrative examples to derive limits on the aspect
ratios of these structures. As a zero-order approximation we can consider the model by
\citet{bus07a}, neglecting any slowing process \citep[such as phase mixing or thermal exchanges, see e.g.][]{sp99,sp02,d09}.
In this case the buoyancy velocity is close to the Alfv\`{e}n velocity and the aspect ratio,
for the cases discussed by \citet{bus07a}, turns out to be in the range 200$-$300, suitable for Li production.
Alternatively, a less efficient transport is obtained if the instabilities can travel only a limited distance
$l$ before interfering destructively (by phase mixing) with previously-detached buoyant structures.
The length covered is $l = \pi \Delta t_A /t$, where $t$ is the available time for the mixing and $\Delta t_A$
is the crossing time of Alfv\`{e}n waves over the star's dimension \citep{sp99,b04}. In such a case,
using the parameters of RGB phases by \citet{bus07a}, one can easily derive that the aspect ratio becomes
very small (of the order of 0.1$-$0.2), suitable to account for the requirements of the carbon isotopic ratios
in red giants and with a correspondingly small velocity. A similarly small velocity is
obtained for the magneto-thermohaline mixing by \citet{d09}. These are only very simplified,
order-of-magnitude estimates, derived just by applying to magnetic buoyancy some of the
considerations by \citet{den10a}. However, they are already sufficient to underline that this kind
of mechanism deserves now to be carefully modeled, as it seems to be able to cover the whole range
of mixing velocities (or aspect ratios) required by virtually all the chemical peculiarities displayed
by red giant stars.

\section{Conclusions}

In this paper we presented revised parametric calculations of
nucleosynthesis in extra-mixing phenomena occurring in low mass
evolved stars, from the phase of the FDU up to the end of the AGB stage.
The calculations were based on a recent revision of the
FRANEC code and on a new extra-mixing code. They incorporated several
physical and numerical improvements, among which new opacities for
carbon-rich envelopes, a new algorithm for the determination of the
convective border position, and a new mass-loss law.

With these tools we explored the coupling of nucleosynthesis and
non-convective mixing in RGB and AGB phases, comparing the results
obtained with the old NACRE compilation of reaction rates with those
descending from the adoption of new recommendations from ADE10 and
ILI10. In particular, the specific choices made for each updated
reaction rate were commented extensively in the Appendix.

It was found that, among the new reaction rates, the ones inducing
the largest variations in the evolution of LM stars are those for
proton captures on $^{14}$N and (to a lesser extent) for proton captures
on $^{17}$O. Similar conclusions were recently obtained in another
context by \citet{magic}. We obtained huge changes in the oxygen isotopic
composition of the envelope, starting from FDU. Their
effects include a drastic shrinking of the mass interval in which
AGB stars can produce oxide grains like those recovered in pristine
meteorites. Most such grains appear to come from stars below 1.5
M$_{\odot}$. Another important revision of previous knowledge was the
discovery that all $^{18}$O-poor grains (those with $^{18}$O/$^{16}$O below 0.0015)
can now be explained, without the "forbidden regions" previously found for
the lowest $^{17}$O/$^{16}$O ratios. Concerning the evolution of AGB stars of low mass and solar
metallicity we suggested that most MS, S stars should have masses
below 1.5 $-$ 1.7 M$_{\odot}$; in this mass interval they would never produce
C-stars, also thanks to $^{12}$C processing in deep mixing.

Constraints coming from observations of CNO nuclei in both RGB and
AGB stars were well reproduced by deep mixing models with the new
rates. In particular, for RGB stars we found that extra-mixing
processing does not need to penetrate to very high temperatures, so
that thermohaline diffusion might be an adequate scenario for
those phases, if the problems recently mentioned in the literature
for its too low diffusion coefficients can be solved.
We then confirmed that deep mixing is required also on the AGB.
In this case, the abundances of $^{26}$Al in presolar grains
indicate that the transport must reach much hotter layers than in
RGB stages. This fact (and the previous consumption of $^3$He) induced us to
conclude that thermohaline diffusion is not a viable explanation
for the transport processes occurring in the latest
evolutionary stages. The existence of a small population of oxide
grains with low $^{17}$O/$^{16}$O and high $^{26}$Al/$^{27}$Al
ratios was suggested to require some further revision of the rates for
$^{26}$Al synthesis at low temperatures,
% probably an increase for the effects of the 93keV resonance on the $^{25}$Mg(p,$\gamma$)$^{26}$Al rate.
probably an increase for the $^{25}$Mg(p,$\gamma$)$^{26}$Al rate.

The longstanding impossibility of accounting for the whole range of
the nitrogen isotopic ratios in SiC grains remains unchanged after our work.
It was argued that no stellar process occurring in low mass stars can solve
this problem and that SiC grains might be contaminated by cosmic ray
spallation processes in the interstellar medium.

\acknowledgements{We are indebted to many colleagues for helping us in the understanding
of the subtleties of low mass star nucleosynthesis. While recognizing the role of
all of them is impossible, we cannot avoid acknowledging the fundamental lessons we learned
from friends like G.J. Wasserburg, R. Gallino, O. Straniero and K.M. Nollett.
%S.P. is grateful to the University of Perugia and to INFN for supporting financially her post-doc fellowship.
M.L. acknowledges the support of the Italian Ministry of Education, University and Research
under Grant No. RBFR082838 (FIRB2008)}

\appendix
\section{The choice of nuclear reaction rates}

\subsection{The reactions ${}^{14}{\rm N}(p,\gamma){}^{15}{\rm O}$ and ${}^{15}{\rm N}(p,\gamma){}^{16}{\rm O}$.}

The ${}^{14}{\rm N}(p,\gamma){}^{15}{\rm O}$ reaction is the slowest
reaction of the CN cycle,  hence the one determining the duration of
the process. The low-energy cross section is dominated by a
resonance at 259 keV. In particular, at the temperatures of interest
here ($T_9 \lesssim 0.06$), the reaction rate mainly depends on the
resonance tail, the Gamow window extending (at most) to around 110
keV. Such a region is only partly covered by experimental data,
spanning energies as low as 70 keV, so the S-factor is very
sensitive to the extrapolating curve and can be severely affected by
the $-504$ keV subthreshold resonance. In the last years, direct
measurements (in both surface and underground laboratories) and
indirect measurements (Doppler shift attenuation, Coulomb
excitation, Asymptotic Normalization Constant) have been performed,
primarily triggered by the discovery that the contribution of the
ground-state transition had been previously overestimated. From
these measurements, ADE10 recommend $S(0)=1.66 \pm 0.12$ keV\,b,
which is about half the NACRE value $S(0)=3.2 \pm 0.8$ keV\,b. This
entails a significantly lower reaction rate in the range of our
interest, as shown in Fig. \ref{f1}, left panel.

The ${}^{15}{\rm N}(p,\gamma){}^{16}{\rm O}$ reaction fixes the leak
rate from  the CN-cycle to the CNO bi-cycle. Besides the direct
capture contribution, the cross section is characterized by the
interference of two $1^-$ resonances at 312 and 964 keV. As for
${}^{14}{\rm N}(p,\gamma){}^{15}{\rm O}$, the largest energy of
interest is $\sim 110$ keV, thus the extrapolation procedure is
critical. In NACRE, a very large S(0) was obtained ($64 \pm 6$
keV\,b)  while ADE10 recommend $36 \pm 6$ keV\,b following the
results from an underground measurement (which does not cover the
Gamow window anyway) and an ANC measurement, yielding an absolute
value of the direct-capture contribution ten times  smaller than
NACRE. Consequently, a 30-40\% smaller reaction rate is now obtained
in the temperature region we are dealing with (Fig. \ref{f1}, right
panel), by using the low-energy S(E)-factor recommended by ADE10.

\subsection{The reactions ${}^{16}{\rm O}(p,\gamma){}^{17}{\rm F}$,  ${}^{17}{\rm O}(p,\gamma){}^{18}{\rm F}$, and ${}^{17}{\rm O}(p,\alpha){}^{14}{\rm N}$.}

The operation of the ${}^{16}{\rm O}(p,\gamma){}^{17}{\rm F}$
reaction contributes to establish the ratios of oxygen isotopes in
red giants and plays therefore a major role in the analysis of
oxygen-rich presolar grains of circumstellar origin. At the energies
of interest for this work, and up to $\sim 120$ keV, the cross section
is determined by the direct capture to the ground and first excited
states of $^{17}$F. Direct measurements exist down to about 100 keV,
so extrapolations are needed. In the previous evaluation by NACRE,
the greatest part of the error was attributed to the model assumed
in the calculation. Improved calculations have now reduced the
uncertainty affecting the S(0) parameter, from 30\% to 7.5\% (see
ADE10 for details), while the recommended value is in substantial
agreement with the one from NACRE. Fig. \ref{f2} (left panel) shows
the resulting reaction rate.

Much like the previously-mentioned reaction, also ${}^{17}{\rm
O}(p,\gamma){}^{18}{\rm F}$ is expected to modify the relative
abundances of oxygen isotopes. In the $E_{cm}\lesssim 120$ keV
energy range, the cross section shows a resonance at 65.1 keV that
strongly influences the reaction rate above $T_9=0.02$. At lower
temperatures, the dominant role is played by direct captures, while
the subthreshold resonance at -3.1 keV does not provide any
significant contribution. The 65.1 keV resonance and the direct
capture term have been the subject of several improved calculations,
leading to a change in the resonance energy (it was 66 keV in NACRE)
and in its strength (which is a factor $\sim 3.7$ smaller than given
in NACRE). Concerning the direct-capture contributions, the revised
calculations yield a $\sim 30\%$ smaller S(E) factor below 120 keV,
which is responsible for the reduced reaction rate below $T_9=0.02$,
as shown in Fig. \ref{f2}, central panel.

The ${}^{17}{\rm O}(p,\alpha){}^{14}{\rm N}$ competes with the
${}^{17}{\rm O}(p,\gamma){}^{18}{\rm F}$  channel, removing $^{17}$O
nuclei and turning the nucleosynthesis path back towards nitrogen.
In contrast with the radiative capture channel, the $-3.1$~keV
subthreshold resonance gives a major contribution to the reaction
rate for $T_9 < 0.02$. Updated calculations for these resonances are
given in both the reviews we are considering. In particular, the
-3.1 keV-state proton width is lowered by a factor $\sim 3$ as
compared to NACRE. No expression for the reaction rate is given in
ADE10; hence we use the reaction rate table in ILI10, extrapolating
down to the $T_9 = 0.004 - 0.01$ temperature range. A remarkably
lower reaction rate is thus obtained (see Fig. \ref{f2}, right
panel).
Regarding the $^{18}$O(p,$\alpha$)$^{18}$N reaction, despite the extensive experimental
investigation in the last years \citep[ADE10,ILI10,][]{laco08,laco10b} has produced a significant
change in the recommended reaction rate (about 30\%) and a reduction
of its uncertainty (about a factor of 8) at low temperatures, its
effect on AGB nucleosynthesis turns out to be negligible.
The updated reaction rate has been included anyway in the present work.

\subsection{The ${}^{20}{\rm Ne}(p,\gamma){}^{21}{\rm Na}$, ${}^{21}{\rm Ne}(p,\gamma){}^{22}{\rm Na}$ and ${}^{22}{\rm Ne}(p,\gamma){}^{23}{\rm Na}$ reactions}

The ${}^{20}{\rm Ne}(p,\gamma){}^{21}{\rm Na}$ reaction is the
starting point of the NeNa  cycle and plays a key role, being the
slowest reaction of the cycle. Inside the Gamow window, extending up
to $\sim 130$ keV, the astrophysical factor is dominated by a
subthreshold resonance laying at about -7 keV. Additional 8
resonances show up at higher energies (up to about 2 MeV). In
ADE10 both the resonance excitation energy and its strength
have been re-evaluated using more recent measurements, resulting in
a small increase of the S(E)-factor, by 2.9\%. Moreover, a new fit
of the energy trend for the S(E)-factor has been provided for the
numerical integration of the reaction rate. The resulting rate is
displayed in Fig. \ref{f3}, left panel, which shows a substantial
agreement with the NACRE one and a slightly reduced uncertainty.

For the ${}^{21}{\rm Ne}(p,\gamma){}^{22}{\rm Na}$ reaction, we
adopt the recommendation by ILI10. They make use of
nuclear-physics input parameters from \citet{ILI01}. Resonances were
measured for $E_{cm}\geq121$ keV, corresponding to the upper edge of
the Gamow window.  Two threshold states, corresponding to resonance
energies of $E_{cm}=17$ and 95 keV, have been considered. While the
latter (for which only an upper limit of the spectroscopic factor is
known) adds a negligible uncertainty to the reaction rate, the
former can influence it at low temperatures (below $T_9 \sim 0.02$).
Though the reaction-rate accuracy has been improved with respect to
NACRE around $T_9 \sim 0.03$, for $T_9 \lesssim 0.02$ the total rate
remains uncertain by several orders of magnitude (see Fig. \ref{f3},
central panel).

The rate of the ${}^{22}{\rm Ne}(p,\gamma){}^{23}{\rm Na}$ reaction
is taken from  ADE10 and is shown in Fig. \ref{f3}, right
panel. The main difference with respect to previous compilations is
the treatment of the threshold states. Concerning the 151-keV
resonance, its spectroscopic factor has been considered as a mean
value rather than an upper limit. Lower energy resonances, whose
existence had been considered as tentative, were disregarded by
ADE10. As a result, the recommended rate is smaller than in
NACRE above $T_9 \sim 0.03$, but remains in good agreement with the
previous indications elsewhere. The error has been instead
significantly reduced.

\subsection{The ${}^{23}{\rm Na}(p,\gamma){}^{24}{\rm Mg}$ and ${}^{23}{\rm Na}(p,\alpha){}^{20}{\rm Ne}$ reactions}

The Gamow window of the ${}^{23}{\rm Na}(p,\gamma){}^{24}{\rm Mg}$
reaction extends up to about 140 keV.  Therefore, the main
contribution to the low-temperature reaction rate is due to the
threshold states. ILI10 mostly make use of the updated
nuclear physics input parameters by \citet{HAL04} in this energy
region, which have been indirectly measured by means of $(^{3}{\rm
He},d)$ transfer reactions. The contribution of direct capture to
the cross section has been taken from the same work. The resulting
reaction rate is compared with the NACRE choice in Fig. \ref{f4}
(left panel). The updated value lays well within the upper and lower
limits set by NACRE, although the uncertainty is greatly reduced.

Concerning the ${}^{23}{\rm Na}(p,\alpha){}^{24}{\rm Mg}$ reaction,
the same  considerations as for the ${}^{23}{\rm
Na}(p,\gamma){}^{20}{\rm Ne}$ can be repeated. The resulting
reaction rate is given in Fig. \ref{f4}, right panel. The new rate
is about 100 times smaller than the one recommended by NACRE at the
lowest temperatures, while the uncertainty is significantly smaller.

\subsection{The reactions ${}^{24}{\rm Mg}(p,\gamma){}^{25}{\rm Al}$, ${}^{25}{\rm Mg}(p,\gamma){}^{26}{\rm Al}_{gs}$, ${}^{25}{\rm Mg}(p,\gamma){}^{26}{\rm Al}_m$, ${}^{26}{\rm Mg}(p,\gamma){}^{27}{\rm Al}$.}

The ${}^{24}{\rm Mg}(p,\gamma){}^{25}{\rm Al}$ reaction influences
the nucleosynthesis  of magnesium and aluminum, which are key
elements in the analysis of ancient meteorites and presolar grains.
No experimental data are available inside the Gamow window (which
extends up to $E_{cm} \sim 150$ keV). The most recent measurements
provide the cross section down to 200 keV (see ILI10 and reference
therein), where 9 resonances occur (up to 2.3 MeV). In the review by
ILI10, the reaction rate is calculated from the these data,
including re-normalizations of the strengths of higher-energy
resonances and updated Q-values. Moreover, with respect to the NACRE
compilation, a more reliable direct capture contribution to the S(E)
factor has been estimated, as well as the contribution of the 214
and 402 keV resonance tails, by means of numerical integrations.
Despite all these improvements, the resulting reaction rate (Fig. \ref{f5}, left panel) is in overall fair agreement with the NACRE
recommendations. It is shown for $T_9>0.01$, but the reaction rate
is there $\sim 10^{-35} {\rm cm}^3 {\rm mol}^{-1} {\rm s}^{-1}$, so
that the ${}^{24}{\rm Mg}(p,\gamma){}^{25}{\rm Al}$ reaction does
not contribute to nucleosynthesis at very low energies.

As for the previous case, the ${}^{25}{\rm Mg}(p,\gamma){}^{26}{\rm
Al}_{gs}$  reaction (where "gs" stands for "Ground State")  is
important for us to understand the magnesium and aluminum
nucleosynthesis in evolved red giants. Data exist in a wide energy
range, starting from 37 keV and extensively overlapping with the
Gamow peak. Concerning the nuclear physics inputs, the differences
of the ILI10 review with respect to NACRE are: (i) the use of
re-normalized strengths for resonances above $E_{cm}=189$ keV; (ii)
the adoption of the more accurate ground-state $\gamma$-ray
branching ratios $f_0$ for the 189, 244 and 292 keV resonances; and
(iii) a new calculation of the contribution of the 37 keV resonance.
The reaction rate is displayed in Fig. \ref{f5}, right panel: the
recommended value is in good agreement with NACRE, but the
uncertainty has been reduced by more than a factor of 2 at the
lowest temperatures that are most important for us.

The same considerations done above apply to the ${}^{25}{\rm
Mg}(p,\gamma){}^{26}{\rm Al}_m$  reaction (where "m" indicates the
isomeric meta-stable state), but in addition the isomeric-state
$\gamma$-ray branching ratio $1-f_0$ has been taken into account.
Once again, while the recommended value by ILI10 is in overall
agreement with NACRE, the uncertainties are reduced by more than a
factor of 2 at the temperatures of AGB stars (Fig. \ref{f6}, left
panel)

For the ${}^{26}{\rm Mg}(p,\gamma){}^{27}{\rm Al}$ reaction, Fig.
\ref{f6} (right panel)  shows a remarkable difference between the
updated reaction rate by ILI10 and the NACRE value, ranging from a
factor of 2 up to 8, while the error has been reduced, especially at
high temperatures. Indeed, a different direct-capture component was
adopted, the strengths of the measured resonances (above $E_{cm}\sim
100$ keV) were re-normalized and the contributions of the
lower-energy resonances at $E_{cm}=53$ keV and $E_{cm}=105$ keV were
re-analyzed. One has to remark that some ambiguity on the assignment
of quantum numbers for these levels still exists (see ILI10 for a
more detailed discussion).

\subsection{The reactions ${}^{26}{\rm Al}(p,\gamma){}^{27}{\rm Si}$, ${}^{27}{\rm Al}(p,\gamma){}^{28}{\rm Si}$, and ${}^{27}{\rm Al}(p,\alpha){}^{24}{\rm Mg}$.}

The ${}^{26}{\rm Al}(p,\gamma){}^{27}{\rm Si}$ reaction is important
for establishing  the equilibrium abundance of the radioactive
isotope $^{26}$Al, playing a major role in presolar grain studies,
$\gamma$-ray astronomy and solar system formation \citep[see
e.g.][for this last point]{busso10,bus03,was06}. Fig. \ref{f7} (left
panel) shows that the updated reaction rate by ILI10 lays within the
fiducial interval by NACRE, but at low temperature the recommended
value is strongly decreased. Also the uncertainty has been greatly
reduced. In the estimate  of the reaction rate, ILI10 introduced a
calculated direct-capture component and the contribution of 19
resonances. In particular, for almost all resonances below $E_{cm}
\sim 500$ keV, the resonance energies were computed from updated
excitation-energy determinations, while improved calculations of the
proton widths of unobserved low-energy resonances were used to
evaluate their contribution to the rate.

For  the  ${}^{27}{\rm Al}(p,\gamma){}^{28}{\rm Si}$ reaction, ILI10
made use  of the nuclear physics input parameters reviewed in
\citet{ILI01}. With respect to the NACRE compilation, the
direct-capture contribution was recalculated, the strengths of the
measured resonances were re-normalized (above $E_{cm} \sim 200$ keV,
to be compared with the upper limit of the Gamow window, $\sim 160$
keV) and the contribution of four sub-threshold resonances were
re-evaluated (the most important are at $E_{cm}=72$ keV and
$E_{cm}=85$ keV). In particular, an improved determination of their
spectroscopic factors was performed, based on the re-analysis of the
${}^{27}{\rm Al}({}^{3}{\rm He},d){}^{28}{\rm Si}$ stripping data.
Consequently, the recommended total rate deviates from NACRE up to a
factor of 20 in the energy range common to AGB stars (Fig. \ref{f7},
central panel), mainly because of improved estimates of
spectroscopic factors for the threshold states.

A total of 91 resonances at energies of $E_{cm} = 72 - 2966$ keV
were taken into  account by ILI10 for calculating the reaction rate
for the process ${}^{27}{\rm Al}(p,\alpha){}^{24}{\rm Mg}$. For the
measured resonances (for $E_{cm} \gtrsim 486$ keV), the same nuclear
physics input parameters considered in \citet{ILI01} were adopted.
The same holds for threshold resonances, whose strengths were
calculated from a re-analysis of the ${}^{27}{\rm Al}({}^{3}{\rm
He},d){}^{28}{\rm Si}$ stripping cross section and from
spectroscopic data in the literature (see ILI01 and references
therein). The improved estimates of spectroscopic factors for the
threshold states, as in the ${}^{27}{\rm Al}(p,\gamma){}^{28}{\rm
Si}$ case, determine an increase of the recommended total rate with
respect to NACRE by a factor of 10 (Fig. \ref{f7}, right panel).

Finally, we recall that (p,$\gamma$) and (p,$\alpha$) captures on
fluorine have  not been addressed by either ADE10 or ILI10. However,
recent measurements \citep{cou,spy2000} only confirmed previous suggestions, so that the
rates do not require updates, and we adopt them from NACRE.

\newpage

\centerline{\Large Figures}

\begin{figure*}
\centering
{\includegraphics[width=\textwidth]{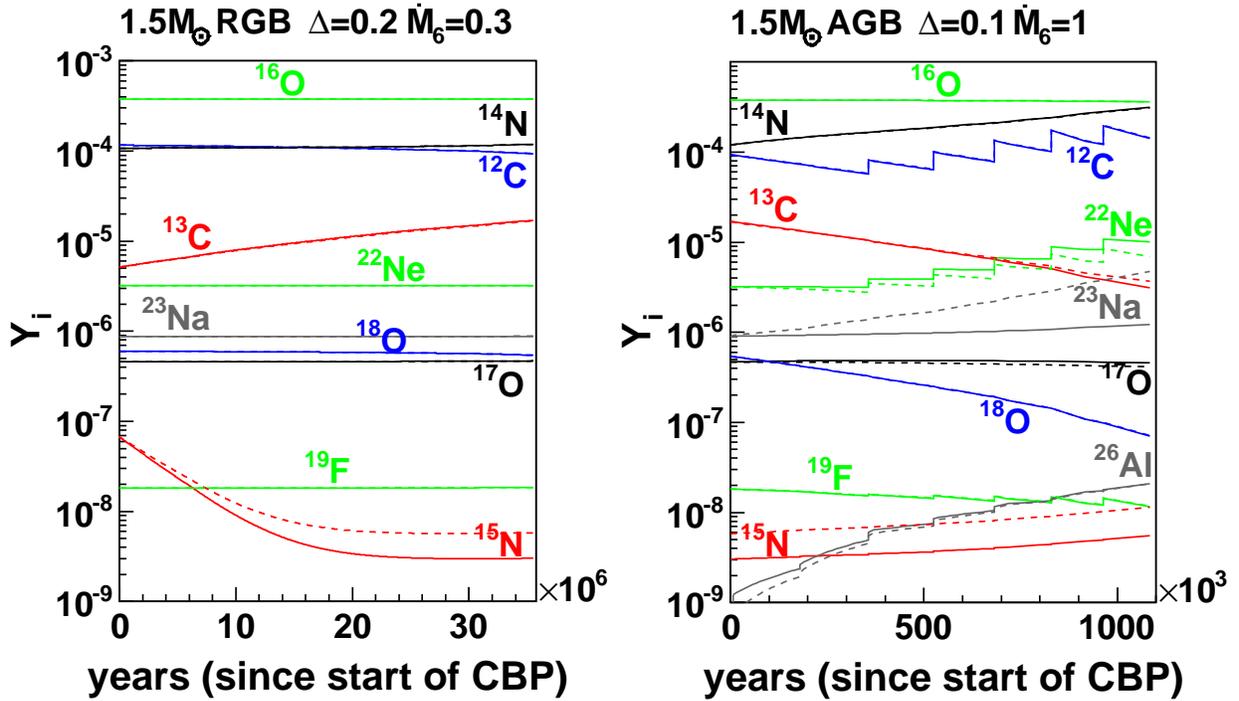}\caption{An example of the evolution in
time  of the abundances by mole in the envelope for representative
isotopes. The RGB and AGB phases of a 1.5 M$_{\odot}$ star of solar
metallicity are shown for a specific choice of the mixing
parameters. (A color version of this figure is available in the online journal.)}}\label{f14}
\end{figure*}

\begin{figure*}
\centering
{\includegraphics[width=\textwidth]{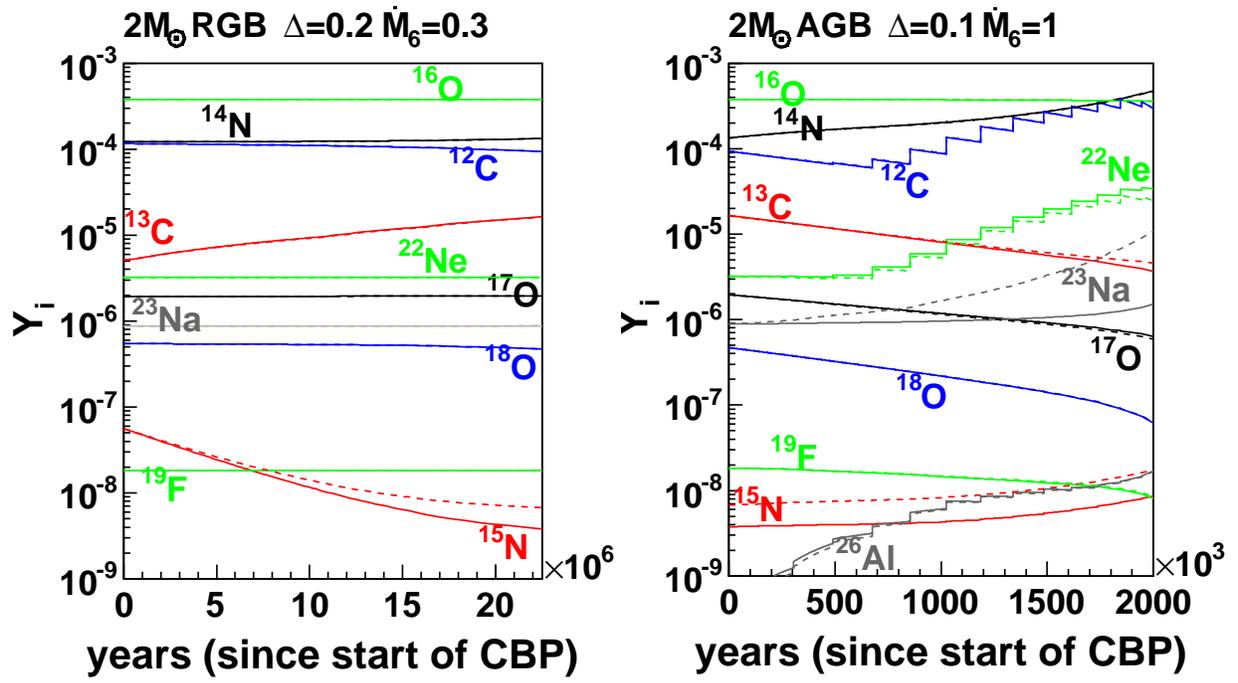}\caption{Same as in the previous
figure, but  for a 2 M$_{\odot}$ model star. (A color version of this figure is available in the online journal.)}} \label{f15}
\end{figure*}

\begin{figure*}
\centering{\includegraphics[width=\textwidth]{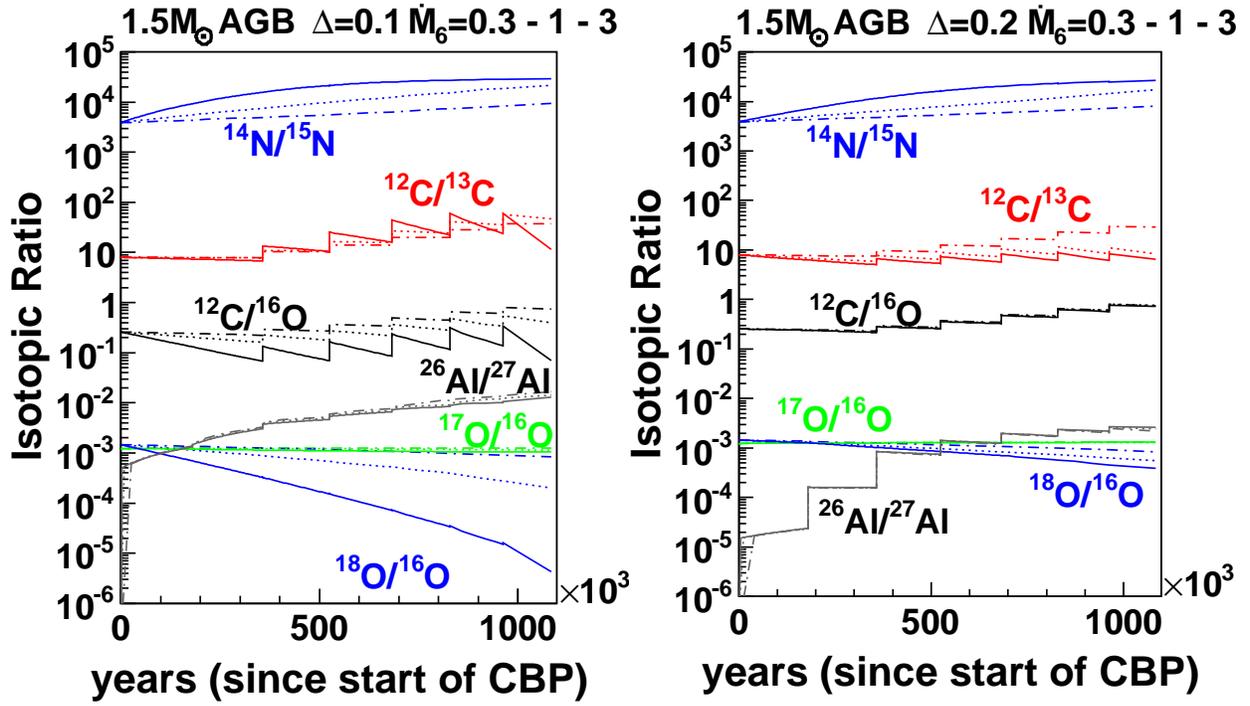}\caption{Evolution in time of the
abundances by mole in the envelope of a 1.5 M$_{\odot}$ star, during
the AGB stages. The plot considers three values of the mixing
parameter $\dot M_6$ and two choices of the temperature parameter
$\Delta$. (A color version of this figure is available in the online journal.)}} \label{f16}
\end{figure*}

\begin{figure*}
\centering{\includegraphics[width=\textwidth]{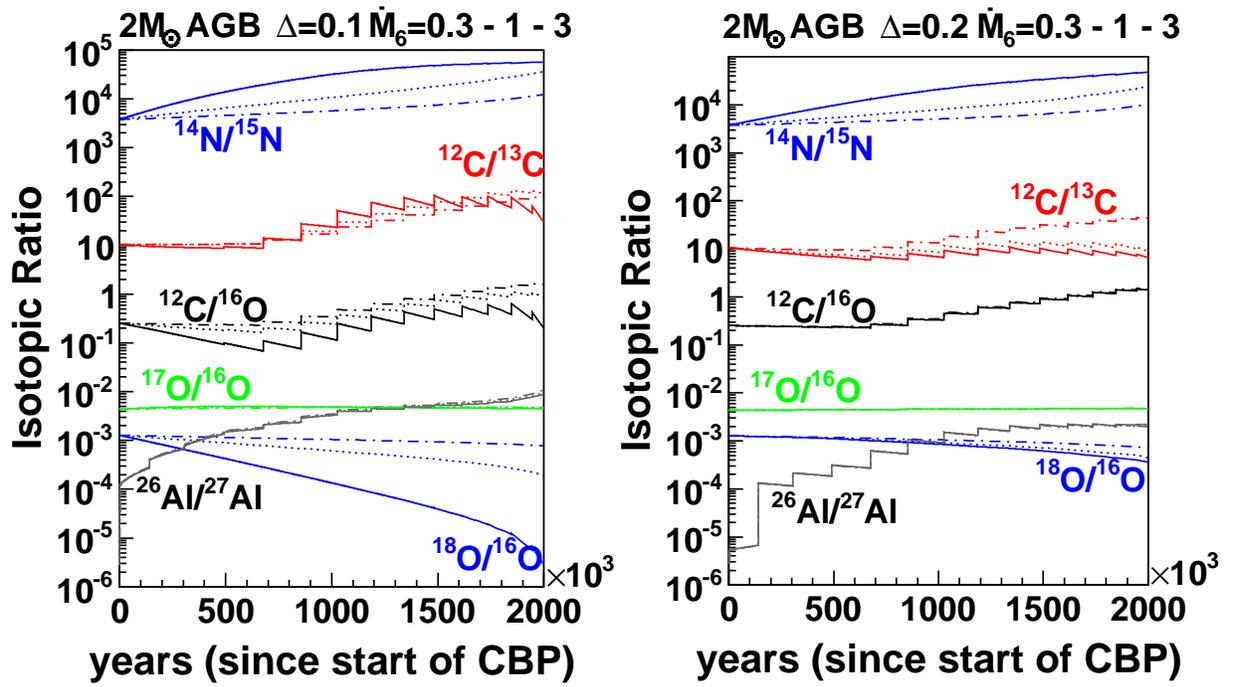}\caption{Same as in the previous
figure, for a stellar model of 2 M$_{\odot}$. (A color version of this figure is available in the online journal.)}} \label{f17}
\end{figure*}

\begin{figure*}
\centering{\includegraphics[width=0.45\textwidth,height=0.35\textheight]{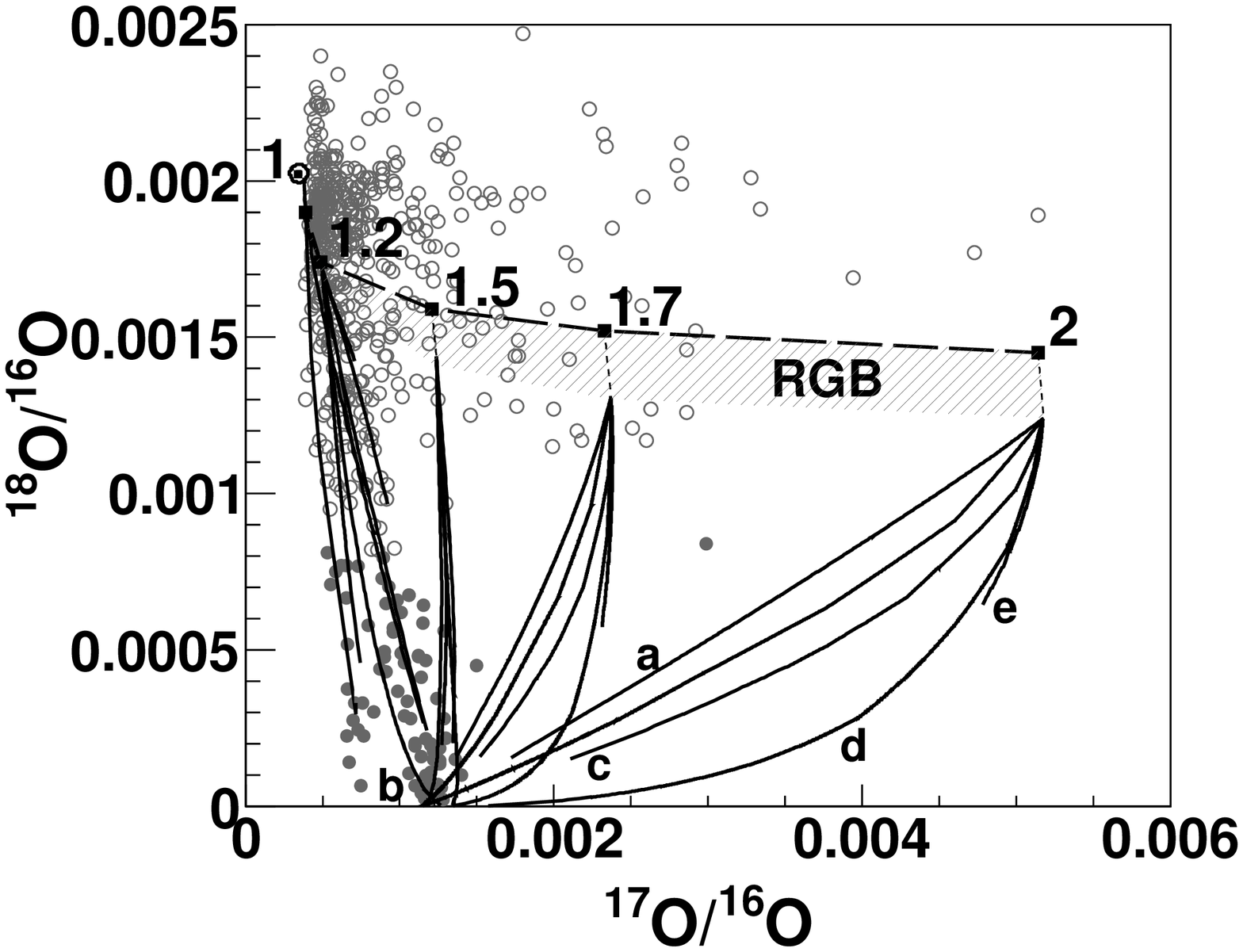}
\includegraphics[width=0.45\textwidth,height=0.35\textheight]{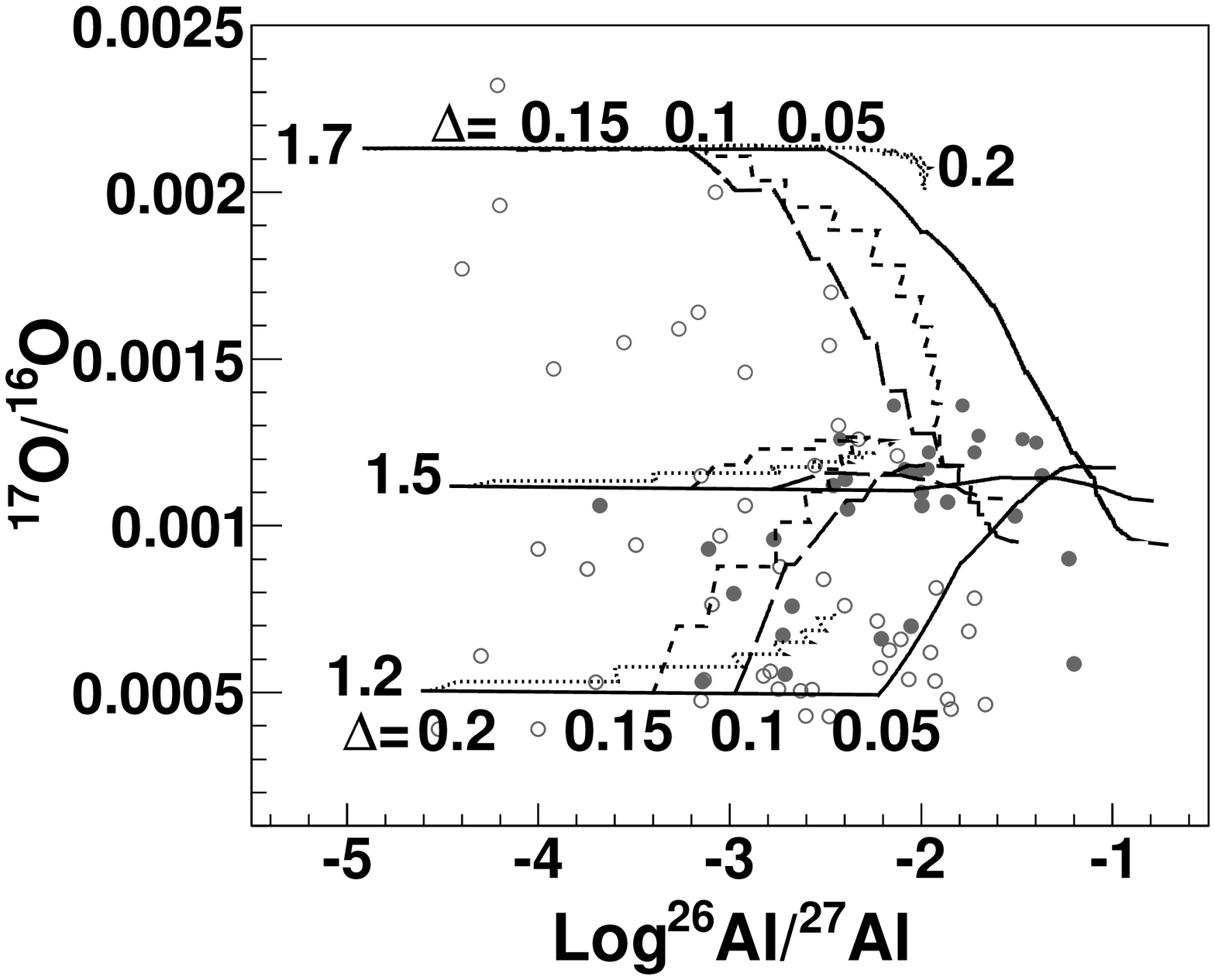}}
\caption{The curves show the oxygen and aluminum isotopic mix in
the  envelopes of our solar-metallicity stars during advanced
evolutionary stages, as compared to presolar oxide grain
compositions (WUSTL Presolar Database $http://presolar.wustl.edu/~pgd/$).
Left panel: FDU is indicated by the heavy, long-dashed
curve; short-dashed lines and the shaded area show the effects of
deep mixing on the RGB, for $\Delta$ = 0.2 and $\dot M_6$ = 0.1.
This is a more moderate choice than in Fig. \ref{f10}. The symbols for
data points are the same as in Fig. \ref{f10}. Continuous lines refer to
model results for extra-mixing on the AGB (all models for a specific mass
are named a), b), c), d), e) from left to right).
They correspond to specific choices of the parameters, namely:
a) $\Delta$ = 0.1, $\dot M_6$ = 1;  b)  $\Delta$ = 0.1,
$\dot M_6$ = 3; c)  $\Delta$ = 0.15, $\dot M_6$ = 1; d)  $\Delta$ =
0.15, $\dot M_6$ = 3; e)  $\Delta$ = 0.2, $\dot M_6$ = 1.
Models for $M \le 1.7 M_{\odot}$ explain essentially all the data for
$^{18}$O-poor grains. In particular, group 2 grains mainly derive
from very low-mass stars (below 1.5 M$_{\odot}$). Right panel: the
$^{17}$O/$^{16}$O ratio as a function of the $^{26}$Al content.
Continuous lines refer to
$\Delta$ = 0.05; long-dashed lines to $\Delta$ = 0.1; short-dashed
lines to $\Delta$ = 0.15; dotted lines to $\Delta$ = 0.20. Model curves
proceed from left to right, converging toward the
equilibrium $^{17}$O/$^{16}$O at the specific temperature of each
model. Most of the data points are accounted for (see text for the
parameter values).}\label{f18}
\end{figure*}

\begin{figure*}
\centering{\includegraphics[width=0.45\textwidth,height=0.35\textheight]{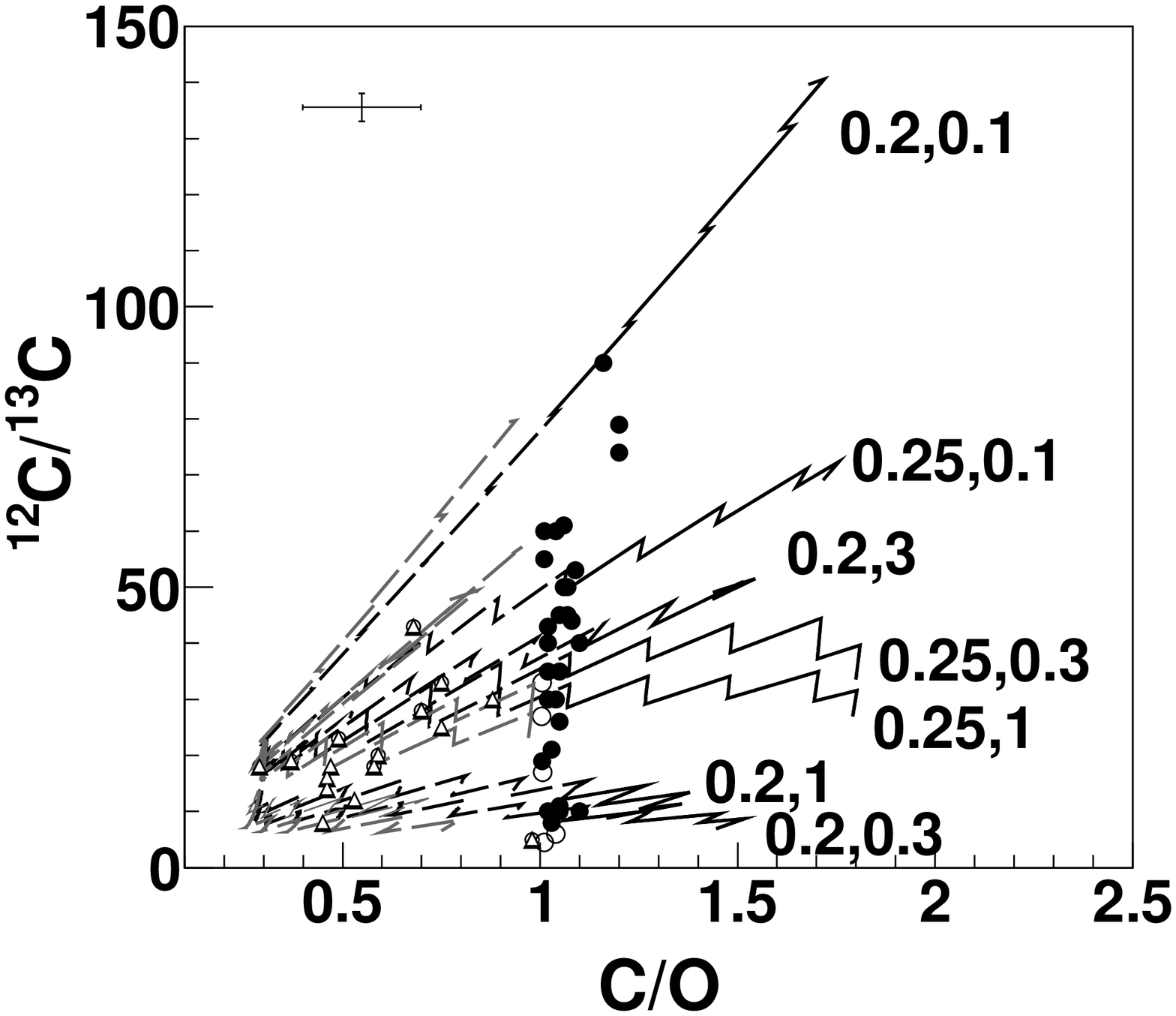}
\includegraphics[width=0.45\textwidth,height=0.35\textheight]{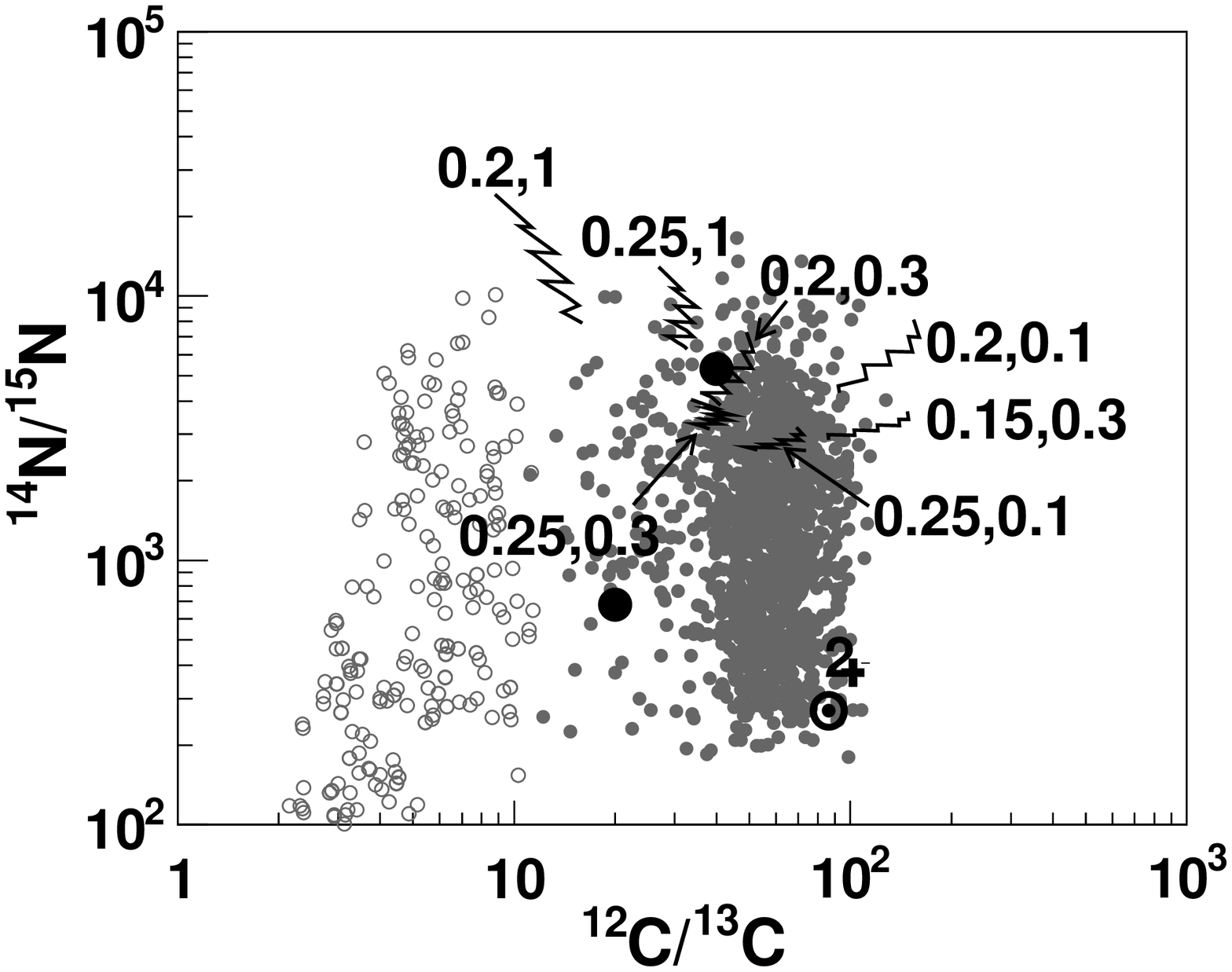}}
\caption{Left panel: observations of C/O and $^{12}$C/$^{13}$C
ratios in AGB stars of types MS, S (white triangles), SC (open circles) and C(N)
(filled circles), as compared to a
few of our curves from models of 1.5 (grey curves) and 2 M$_{\odot}$ (black curves) at solar
metallicity, with extra-mixing and TDU. Continuous lines refer to
the C-rich phases, dashed lines to the O-rich ones. The labels
indicate the choices for $\Delta$ and $\dot M_6$. The area of the
data is well covered by the models, indicating that the parameter
choices should be rather typical of real AGB stars. Right panel:
The $^{14}$N/$^{15}$N ratios of SiC grains recovered from pristine
meteorites, as a function of their $^{12}$C/$^{13}$C ratios. Open
symbols represent the A+B grains, grey symbols the so-called
"mainstream" ones \citep[][WUSTL Presolar Database $http://presolar.wustl.edu/~pgd/$]{zin1,zin2}. Model curves are from a
2 M$_{\odot}$ star of solar metallicity; only the final C-rich phases
are plotted. Again, the model parameters are indicated. The range of
carbon isotopic ratios of mainstream grains and of some A+B grains
is reproduced, but our $^{14}$N/$^{15}$N ratios are always much
larger than solar. The full dots represent stellar measurements by
\citet{wan} in C-rich circumstellar envelopes and confirm that high
values are typical of evolved stars. See text for comments. The two symbols $\jupiter$ and  $\odot$
indicate the values of the nitrogen isotopic ratios found in Jupiter, which we use as initial abundance in our models,
 and in the Sun.}
\label{f19}
\end{figure*}

\begin{figure*}
\centering {\includegraphics[height=6.3cm]{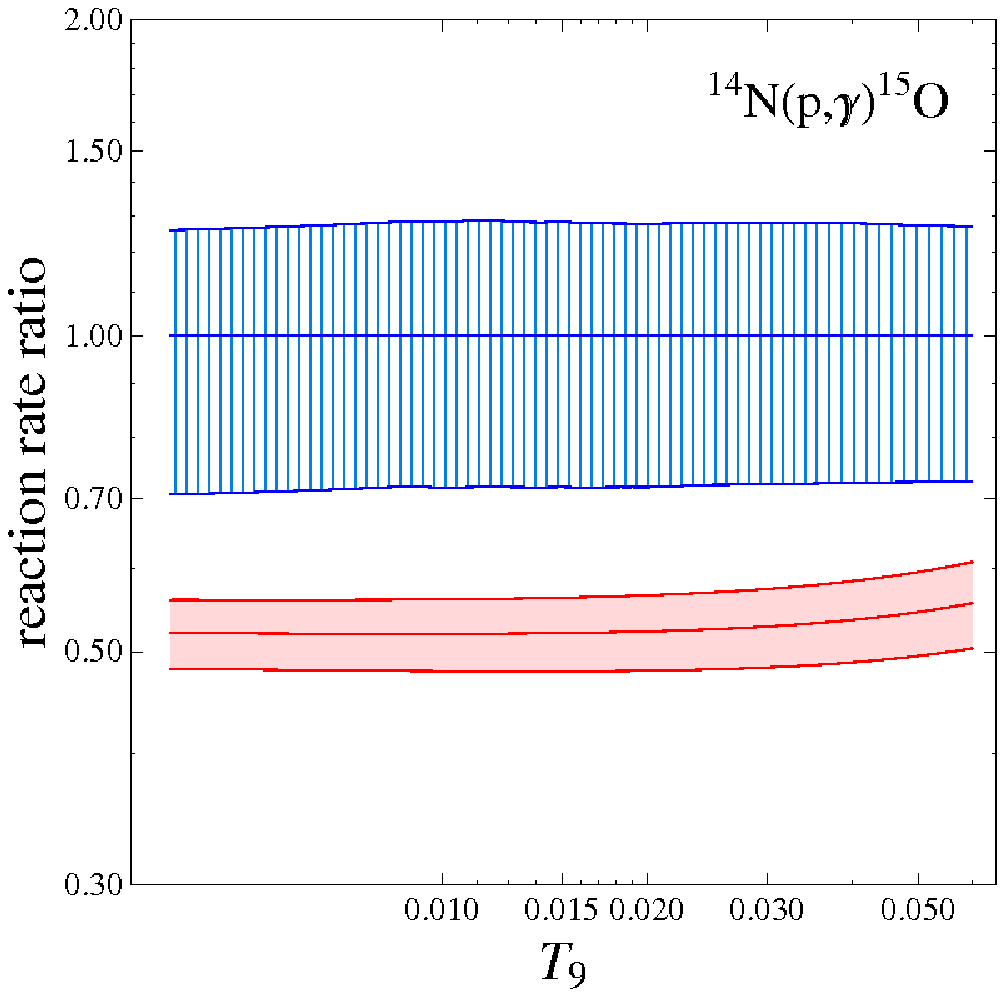}
\includegraphics[height=6.1cm]{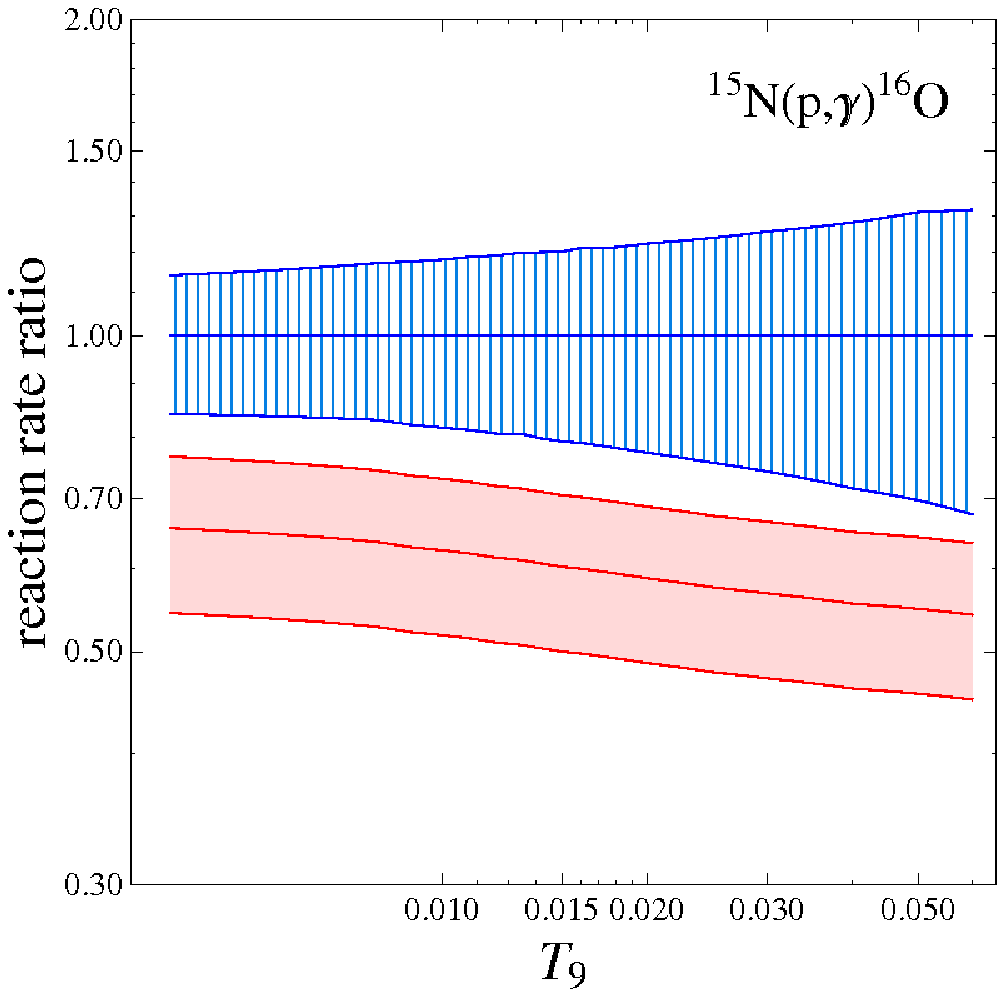}}
\caption{Left: a comparison of the NACRE rate for the reaction
${}^{14}{\rm N}(p,\gamma){}^{15}{\rm O}$  (hatched blue band) with the
updated one from ADE10 (light-red filled band). The solid lines mark the recommended value,
together with lower and upper limits (blue for NACRE, red for
ADE10). The NACRE rate is adopted as a normalization (i.e. it is
taken equal to 1 over the whole temperature region). Right: the same
comparison for the reaction ${}^{15}{\rm N}(p,\gamma){}^{16}{\rm O}$.
 (A color version of this figure is available in the online journal.)} \label{f1}
\end{figure*}

\begin{figure*}
\centering {\includegraphics[height=5.3cm]{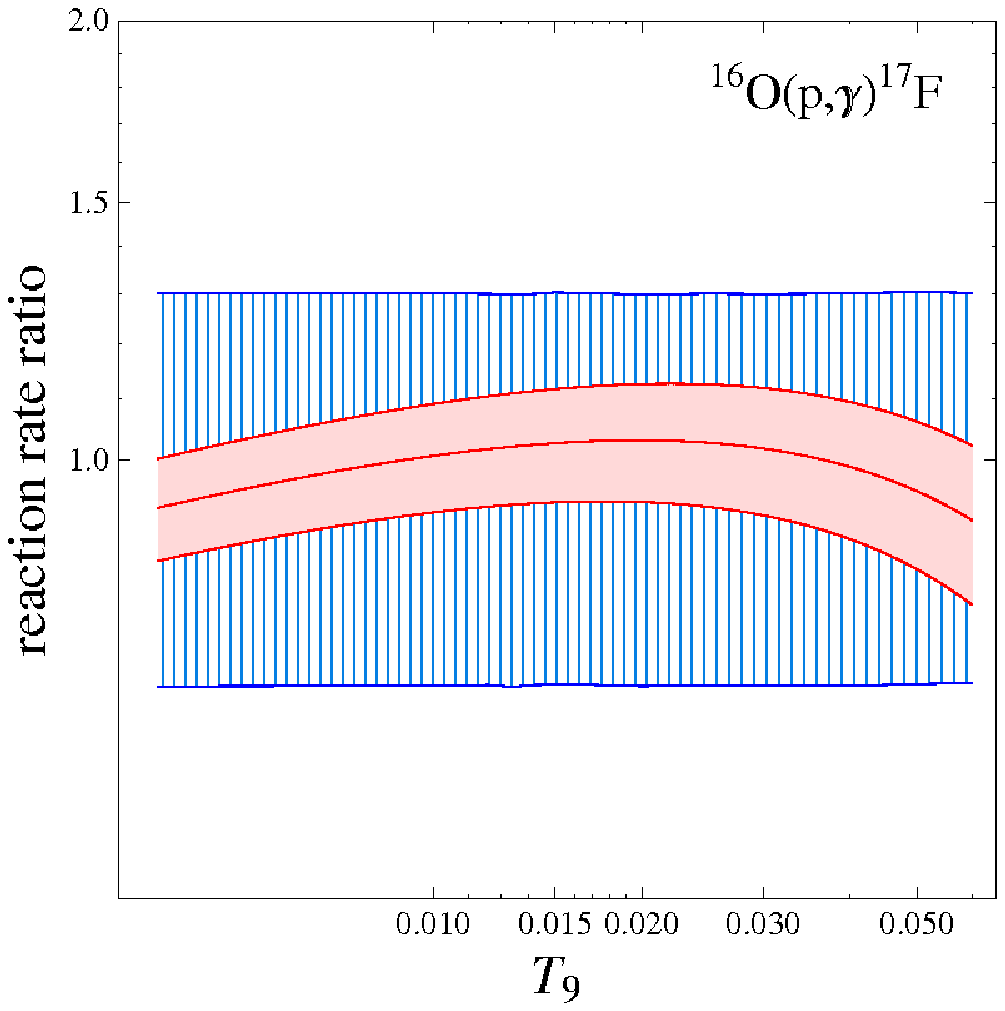}
\includegraphics[height=5.2cm]{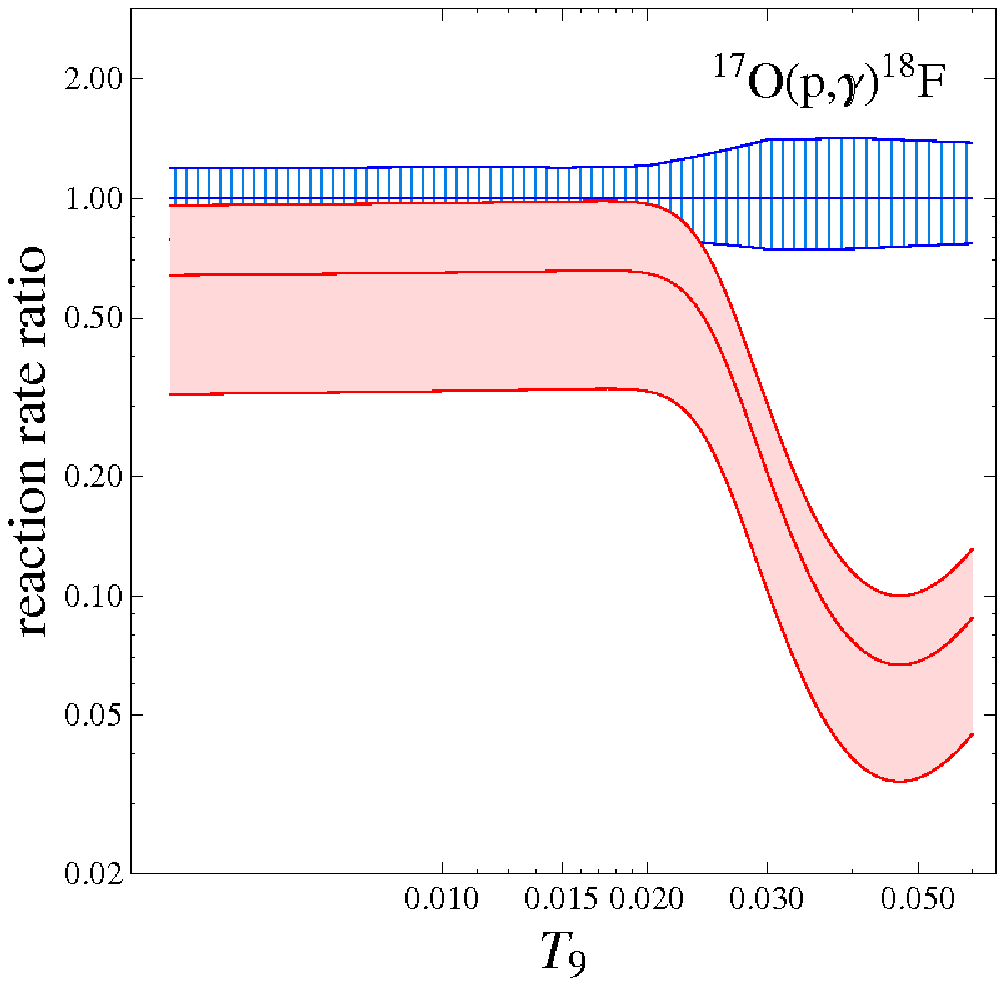}
\includegraphics[height=5.2cm]{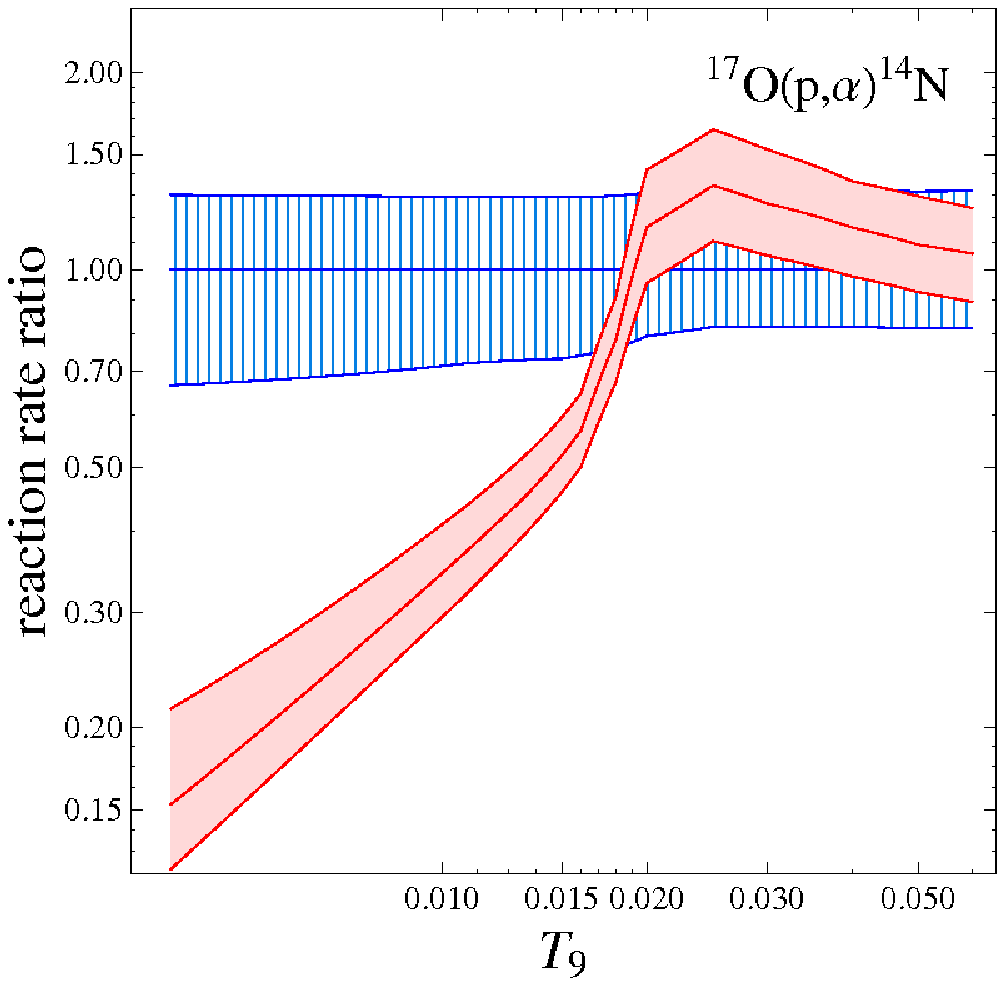}
} \caption{Same comparison as in Fig. 13, for the reactions
${}^{16}{\rm O}(p,\gamma){}^{17}{\rm F}$, ${}^{17}{\rm
O}(p,\gamma){}^{18}{\rm F}$ and ${}^{17}{\rm O}(p,\alpha){}^{14}{\rm
N}$. (A color version of this figure is available in the online journal.)} \label{f2}
\end{figure*}

\begin{figure*}
\centering
{\includegraphics[height=5.3cm]{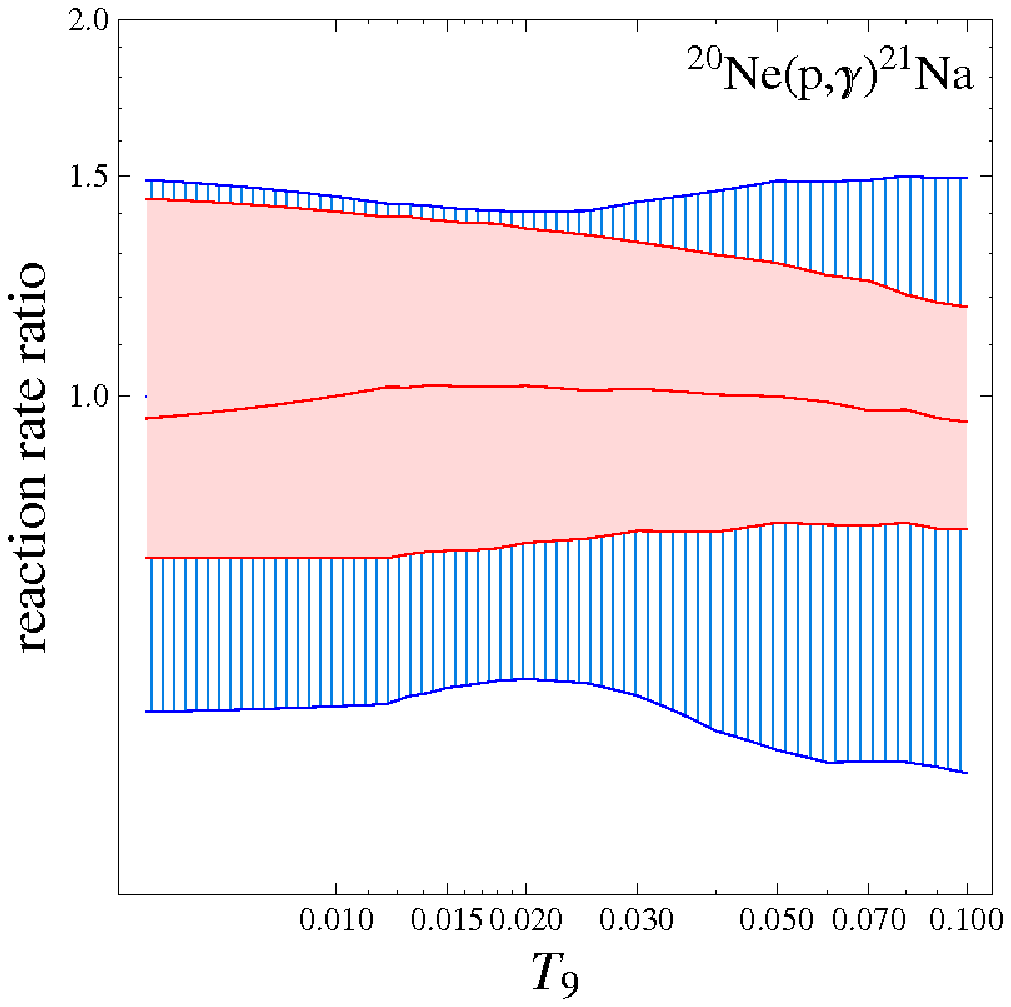}
\includegraphics[height=5.3cm]{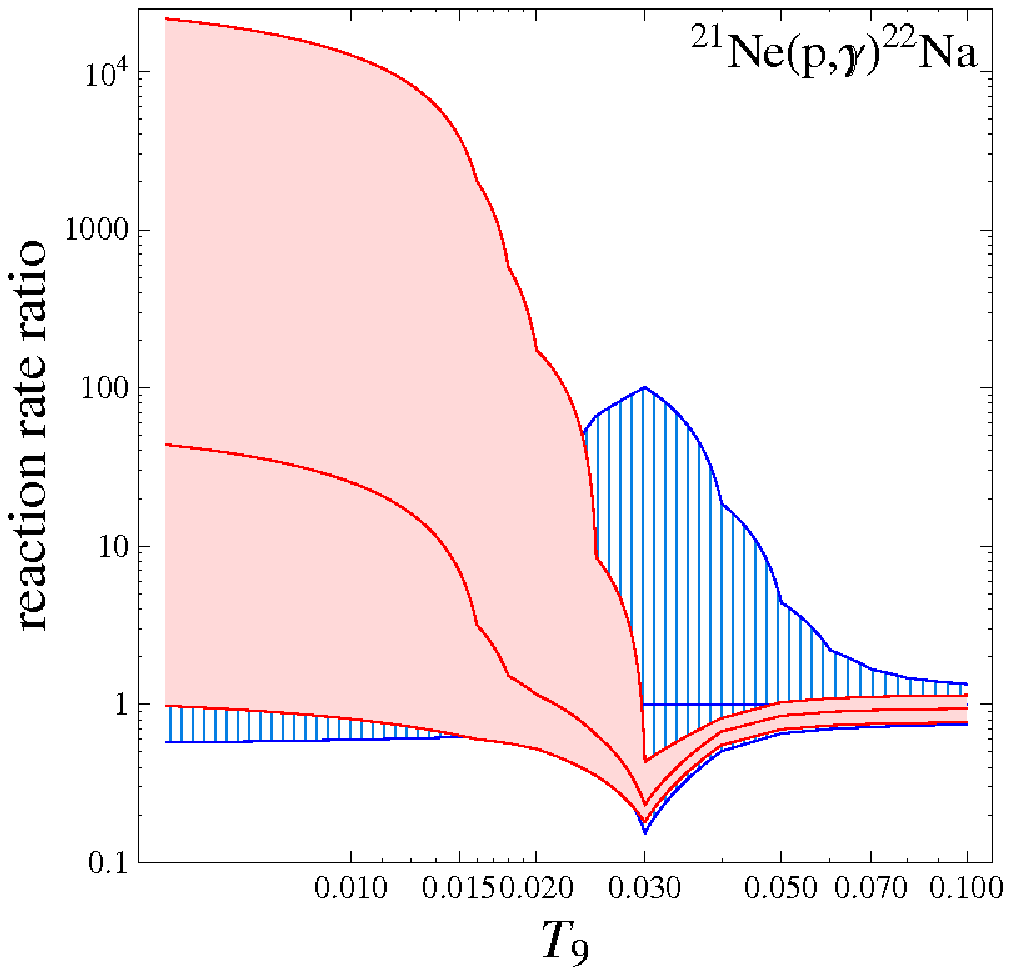}
\includegraphics[height=5.3cm]{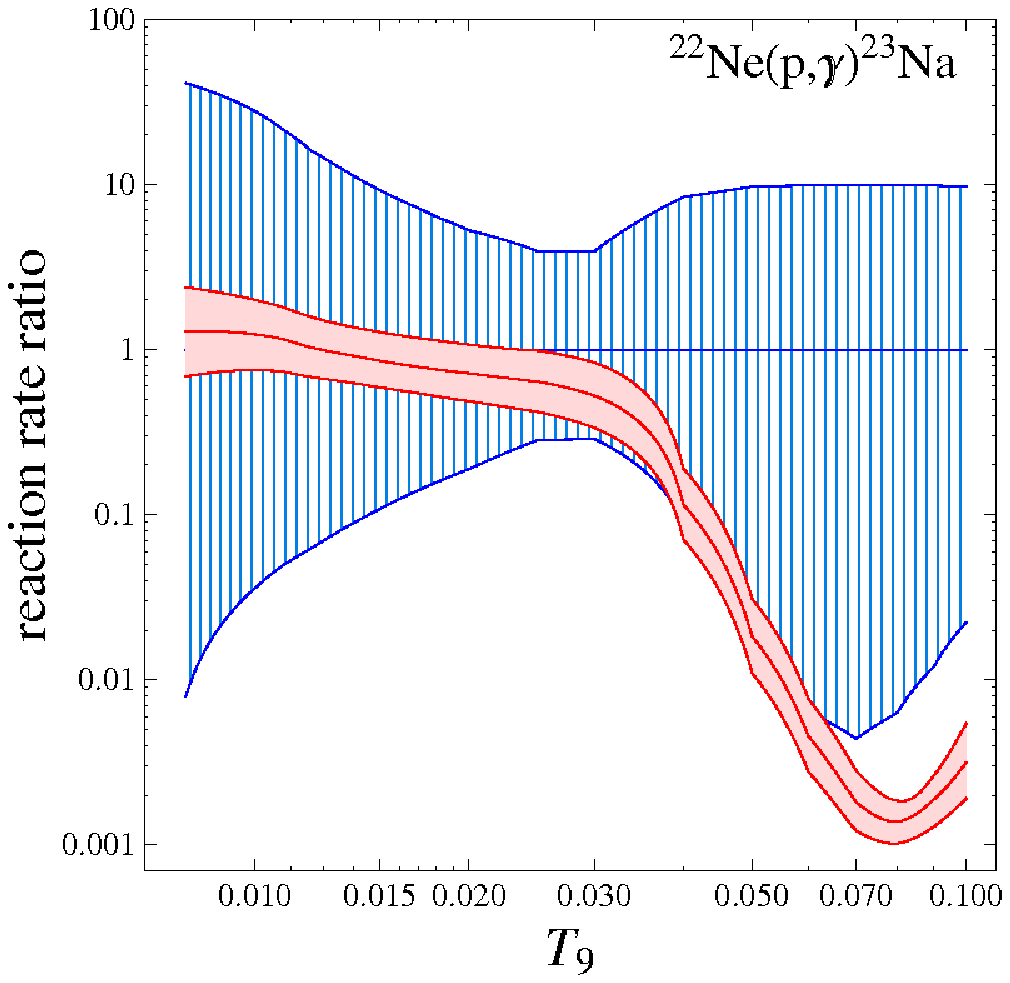}}
\caption{Same comparison as in Fig. 13, for the reactions
${}^{20}{\rm Ne}(p,\gamma){}^{21}{\rm Na}$, ${}^{21}{\rm
Ne}(p,\gamma){}^{22}{\rm Na}$ and ${}^{22}{\rm
Ne}(p,\alpha){}^{23}{\rm Na}$. (A color version of this figure is available in the online journal.)} \label{f3}
\end{figure*}

\begin{figure*}
\centering {\includegraphics[height=6.3cm]{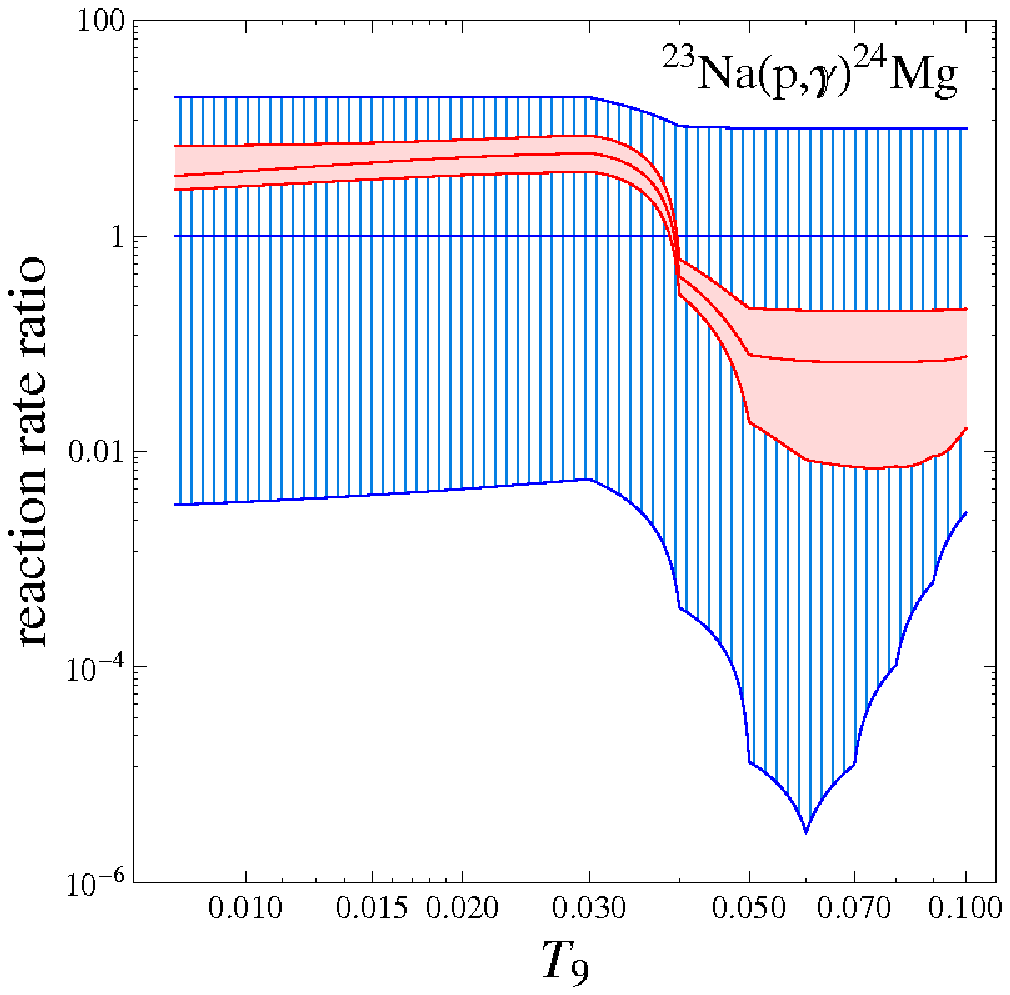}
\includegraphics[height=6.3cm]{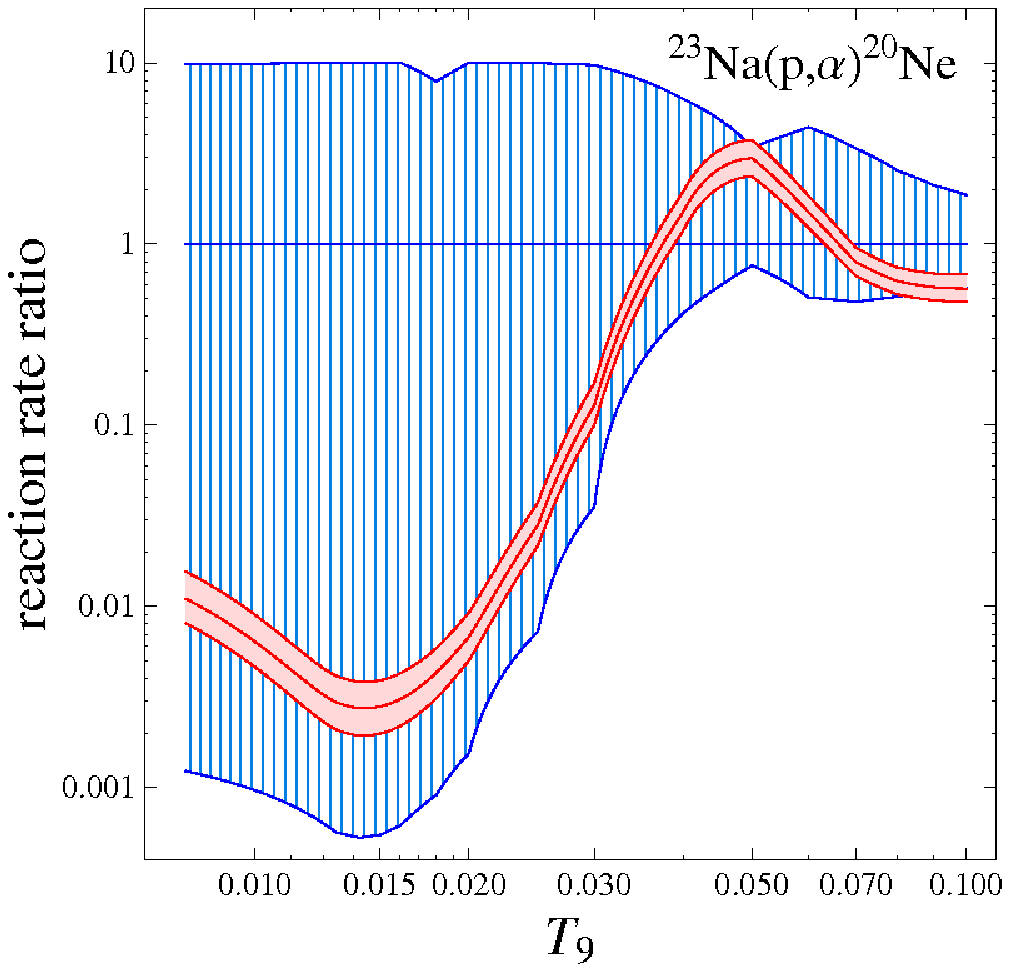}
} \caption{Same comparison as in Fig. 13, for the reactions
${}^{23}{\rm Na}(p,\gamma){}^{24}{\rm Mg}$ and ${}^{23}{\rm
Na}(p,\alpha){}^{20}{\rm Ne}$. (A color version of this figure is available in the online journal.)} \label{f4}
\end{figure*}

\begin{figure*}
\centering {\includegraphics[height=6.3cm]{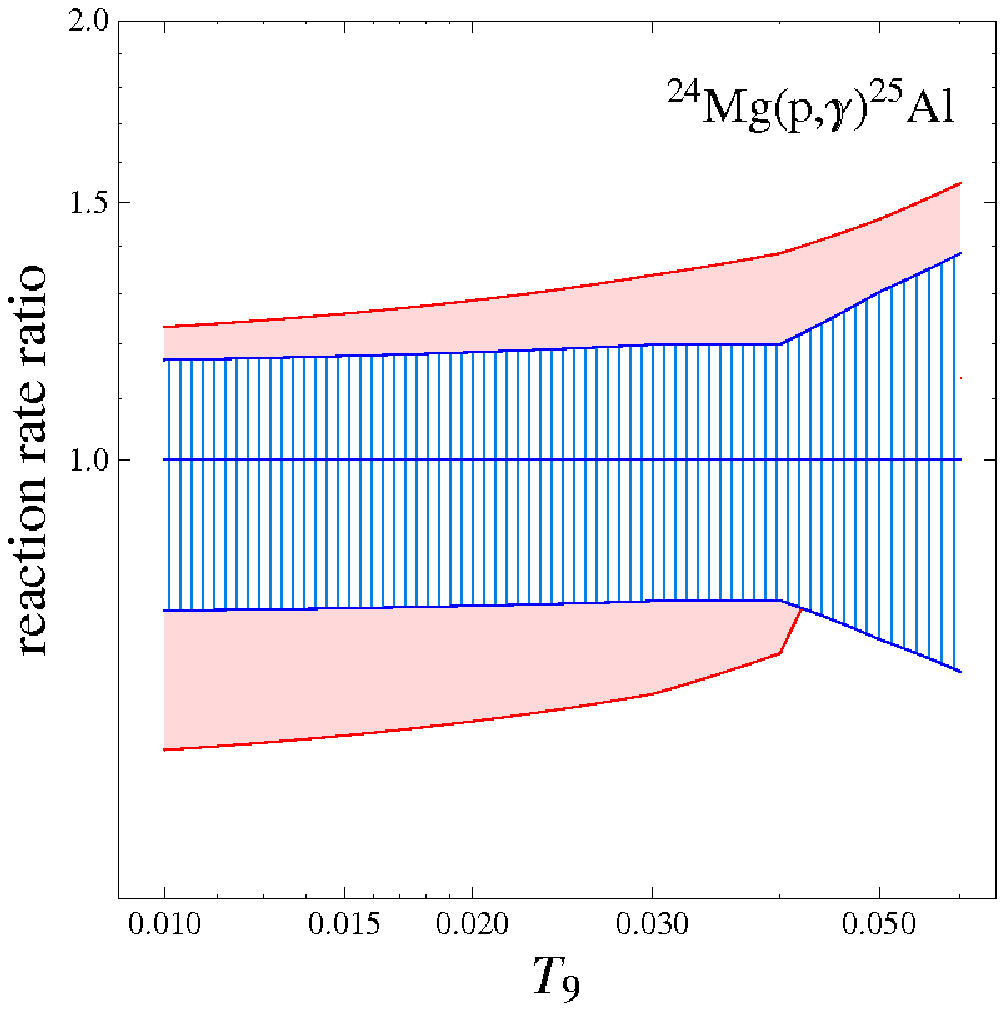}
\includegraphics[height=6.3cm]{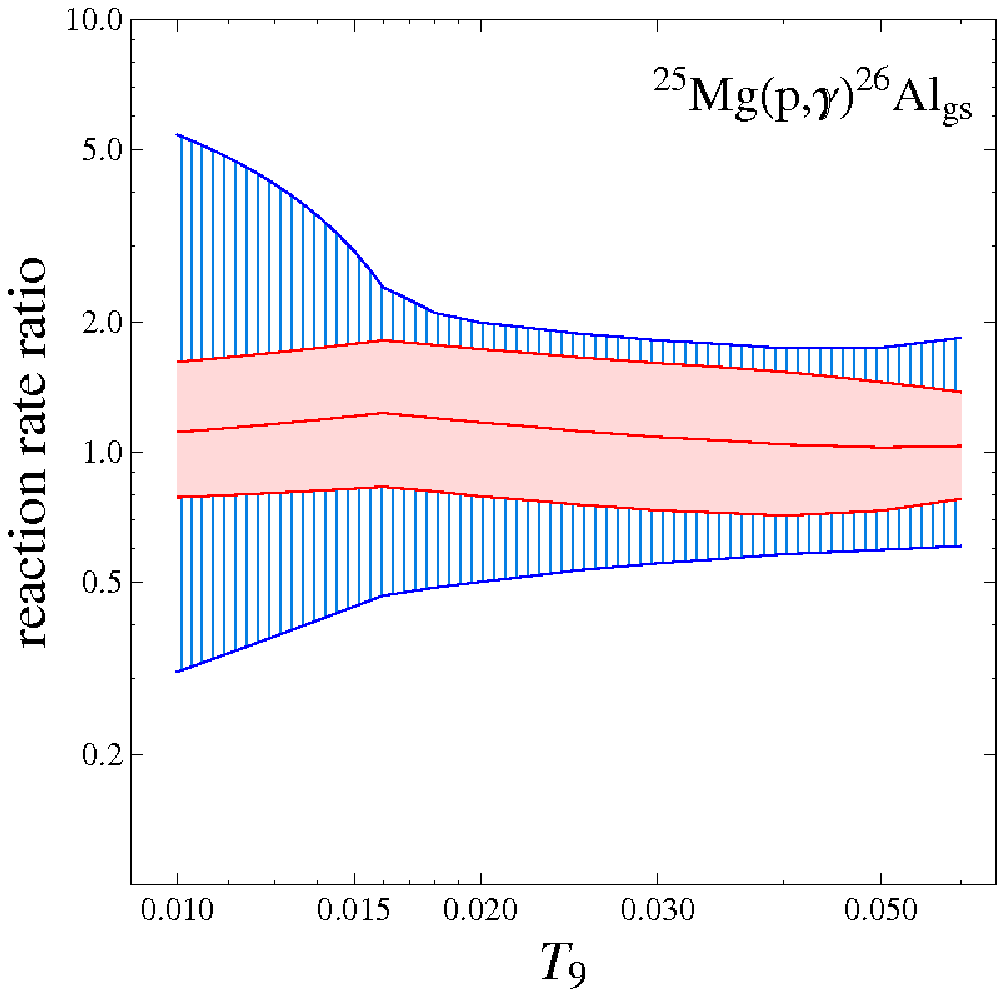}
} \caption{Same comparison as in Fig. 13, for the reactions
${}^{24}{\rm Mg}(p,\gamma){}^{25}{\rm Al}$, and ${}^{25}{\rm
Mg}(p,\gamma){}^{26}{\rm Al}_{gs}$. (A color version of this figure is available in the online journal.)} \label{f5}
%\end{figure*}
%\begin{figure*}
\centering {\includegraphics[height=6.3cm]{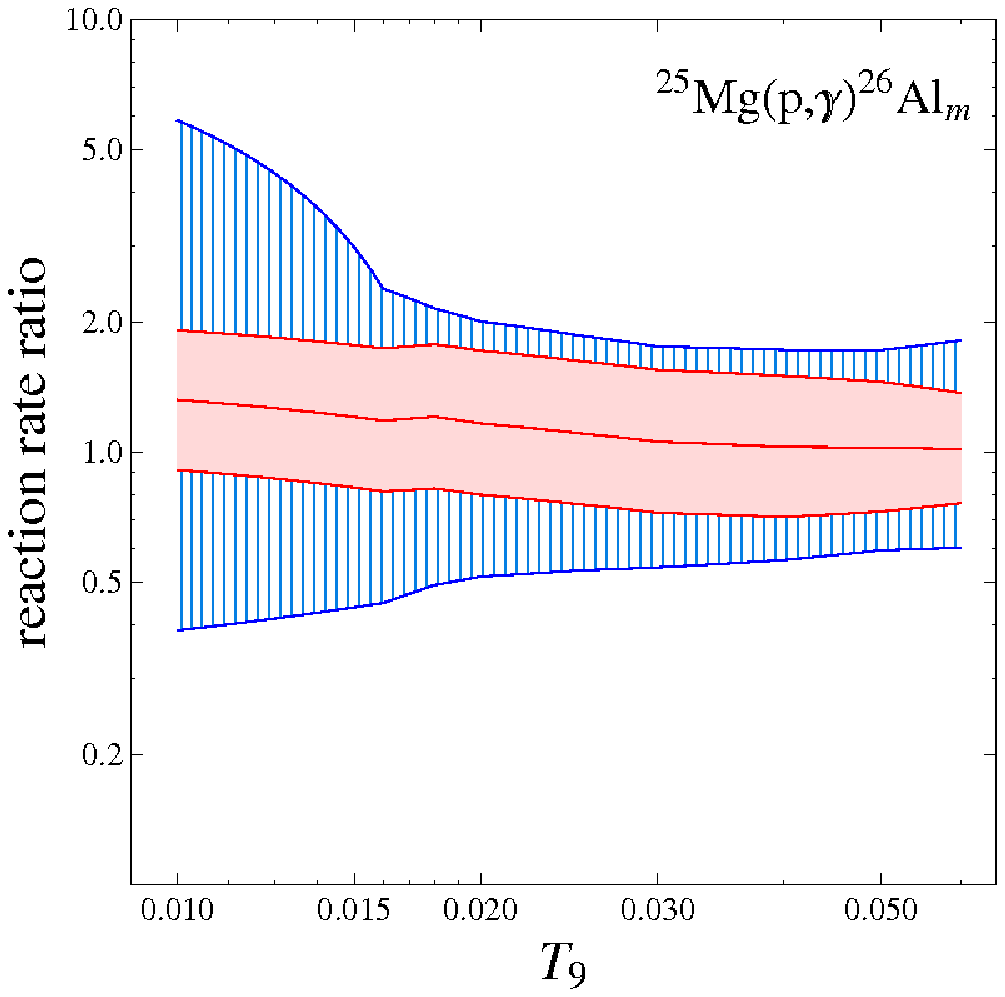}
\includegraphics[height=6.0cm]{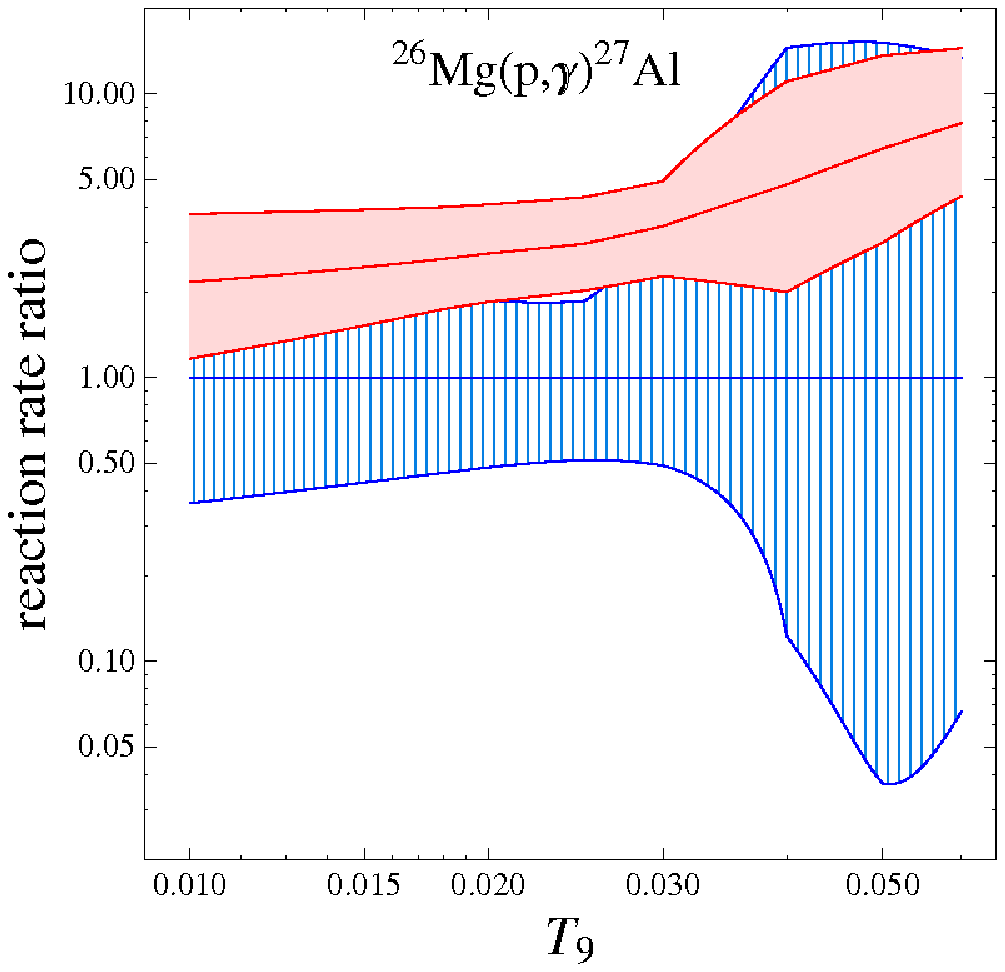}
} \caption{Same comparison as in Fig. 13, for the ${}^{25}{\rm
Mg}(p,\gamma){}^{26}{\rm Al}_m$ and ${}^{26}{\rm
Mg}(p,\gamma){}^{27}{\rm Al}$ reactions. (A color version of this figure is available in the online journal.)} \label{f6}
\end{figure*}
%\end{document}
\begin{figure*}
\centering {\includegraphics[height=5.3cm]{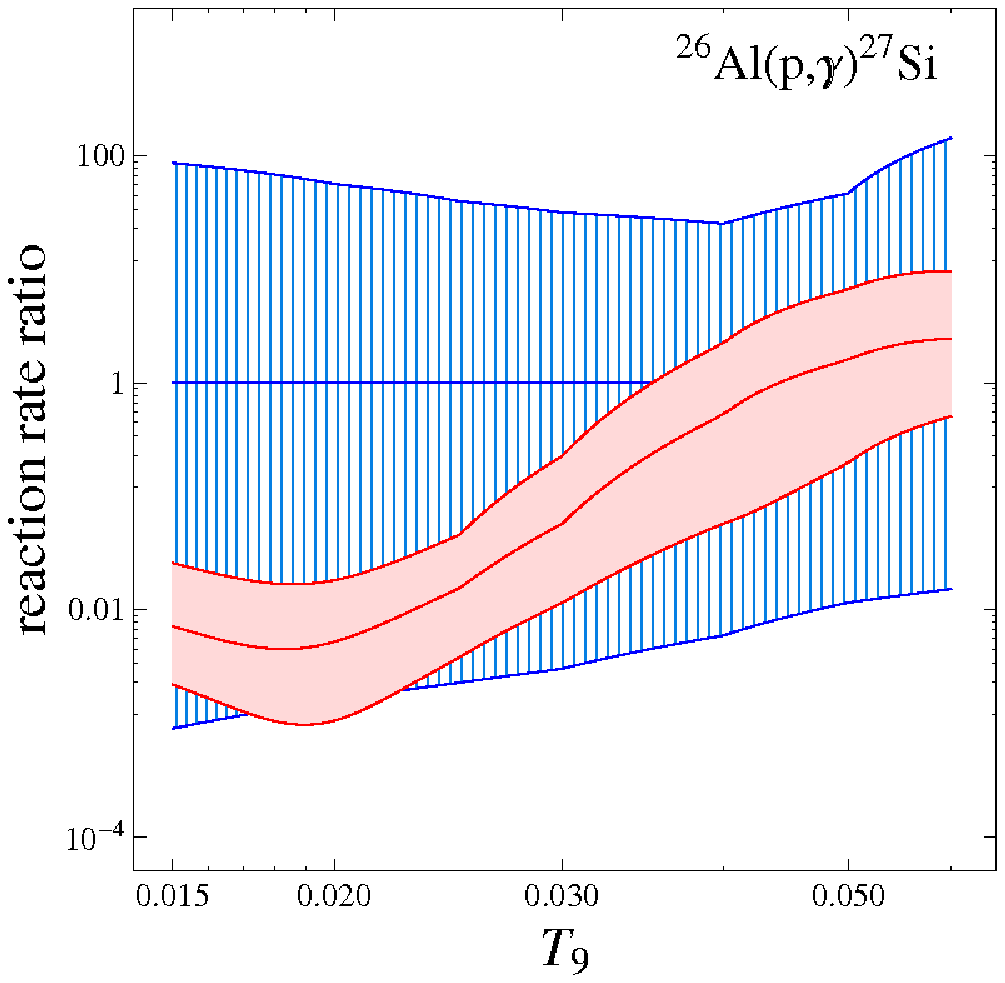}
\includegraphics[height=5.1cm]{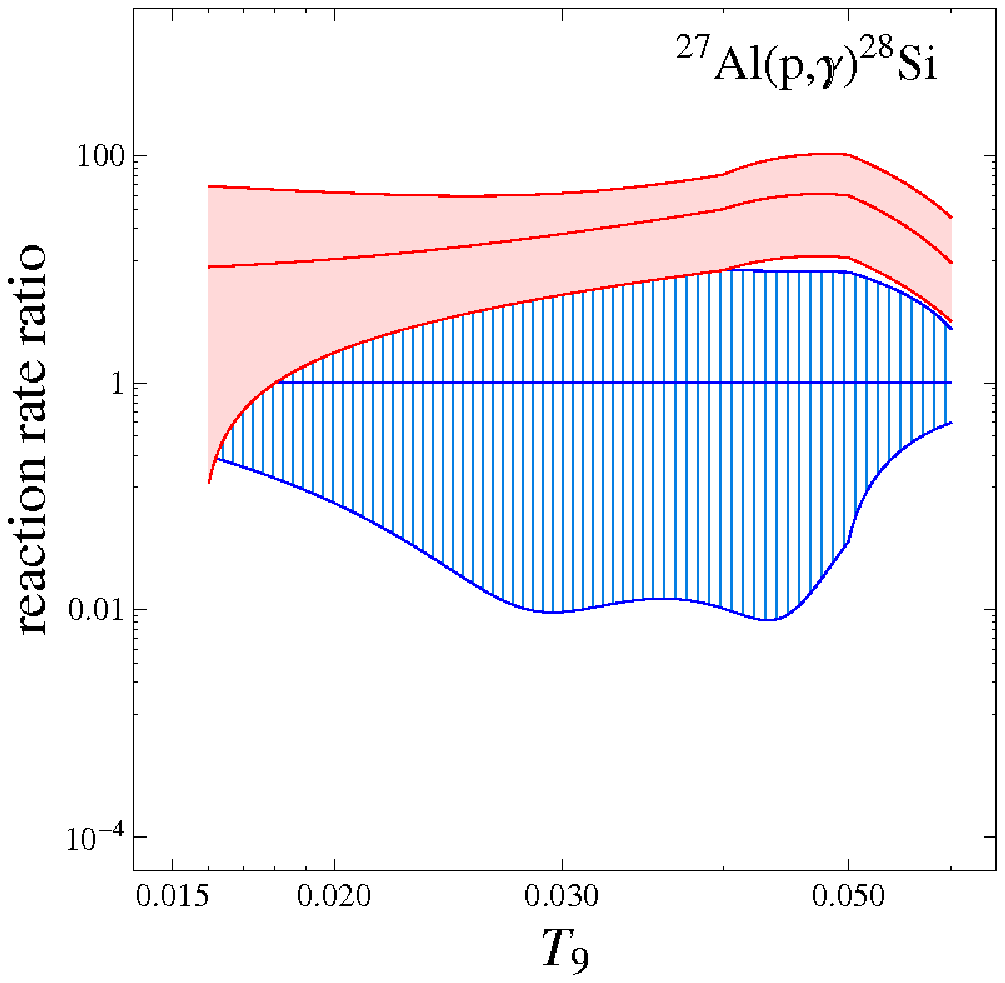}
\includegraphics[height=5.1cm]{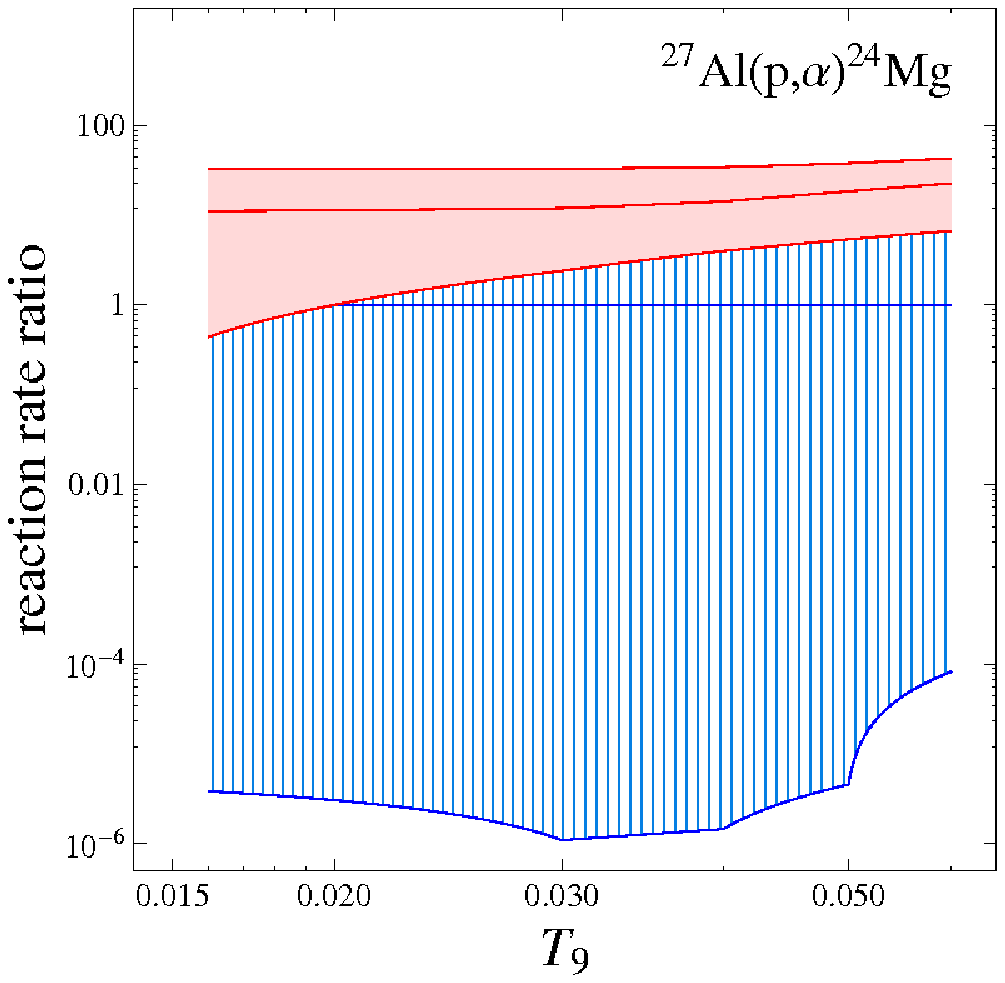}}
\caption{Same comparison as in Fig. 13, for the reactions
${}^{26}{\rm Al}(p,\gamma){}^{27}{\rm Si}$, ${}^{27}{\rm
Al}(p,\gamma){}^{28}{\rm Si}$ and ${}^{27}{\rm
Al}(p,\alpha){}^{24}{\rm Mg}$. (A color version of this figure is available in the online journal.)} \label{f7}
\end{figure*}

\end{document}